\documentclass[journal]{IEEEtran}
\ifCLASSINFOpdf
\else
\fi
\usepackage{cite,array}
\usepackage{amsmath,amssymb,amsfonts,ntheorem}
\usepackage{algorithm}
\usepackage{setspace}
\usepackage{graphicx}
\usepackage{textcomp}
\usepackage{xcolor}
\usepackage{diagbox}
\usepackage{stfloats}
\usepackage[numbers,sort&compress]{natbib}
\usepackage{xcolor}
\usepackage{epstopdf}
\usepackage{algorithmic}
\usepackage{multirow}
\usepackage{booktabs}
\usepackage{subfigure}
\usepackage{epstopdf}
\usepackage{bbm}
\usepackage{makecell}
\theoremheaderfont{\bfseries\upshape}
\theoremseparator{.}
\theoremstyle{plain}
\newtheorem{theorem}{Theorem}
\newtheorem{definition}{Definition}

\def\BibTeX{{\rm B\kern-.05em{\sc i\kern-.025em b}\kern-.08em
    T\kern-.1667em\lower.7ex\hbox{E}\kern-.125emX}}
\begin{document}
\title{Cap the Gap: Solving the Egoistic Dilemma under the Transaction Fee-Incentive Bitcoin}
\author{Hongwei~Shi,
        Shengling~Wang,~\IEEEmembership{Senior Member,~IEEE,}
        Qin~Hu,
        Xiuzhen~Cheng,~\IEEEmembership{Fellow,~IEEE,}
        and~Jianhui~Huang
\thanks{Hongwei Shi and Shengling Wang (Corresponding author) are with the School of Artificial
Intelligence, Beijing Normal University, Beijing, China. E-mail: hongweishi@mail.bnu.edu.cn and wangshengling@bnu.edu.cn.}
\thanks{Qin Hu is with the Department of Computer and Information Science, Indiana University - Purdue University Indianapolis, IN, USA.E-mail:qinhu@iu.edu}
\thanks{Xiuzhen Cheng is with the Department of Computer Science, The George Washington University, Washington DC, USA. E-mail: cheng@gwu.edu}
\thanks{Jianhui Huang is with the Institute of Computing Technology, Chinese Academy
of Sciences, Beijing, China. E-mail: huangjianhui@ict.ac.cn.}
}
\maketitle
\begin{abstract}
Bitcoin has witnessed a prevailing transition that  employing  transaction fees paid by users rather than subsidy assigned by the system as the main incentive for mining.  The adjustability of reward in the transaction fee-incentive regime makes room for the {\it mining gap}, a period of time in which miners turn mining rigs off  until transaction fees are sufficient.  Obviously, the
mining gap aggressively decreases the transaction throughput, weakens the security of Bitcoin, and is further extended by the selfishness of rational users who prone to provide low transaction fees, acting as free-riders. The phenomena of mining gap  and free-riding  trap Bitcoin system into the {\it egoistic dilemma} which is a challenging problem since it involves games  not only between users and miners bilaterally, but also among
miners and users internally. Hence, in this paper, we  first derive the mathematical characteristics of the interplay among users (miners), where the property of  {\it strategic complementarity} of their actions are analyzed.
This enables us to reasonably untangle the antagonism among the  homogenous players,  making all the users (miners) act as a whole to game with their adversary.  Based on this,  an incentive mechanism leveraging  the zero-determinant (ZD) theory is designed  to arm the user-side for inducing the miner-side to  power on its rigs early. Our incentive mechanism is featured by {\it sustained ability}, since the user-side can drive the miner-side to be an ``early bird" without any additional payment in the long run, and {\it fairness}, because even the dominant user-side cannot squeeze miner-side financially. To the best of our knowledge, this paper is the first  work to cap the mining gap and solve the egoistic dilemma under the transaction fee-incentive Bitcoin. Both theoretical analyses and numerical simulations demonstrate the effectiveness of our proposed mechanism.
\end{abstract}

\IEEEpeerreviewmaketitle

\section{Introduction}\label{Intro}
\IEEEPARstart{B}{itcoin} is immune to distrustful entities and can expedite confidence by a {\it mining}-based consensus mechanism without a third party, heralding a new era in digital cryptocurrency since introduced in 2008 \cite{bitcoin,queue,fee}. As incentives for mining, each consensus  participant, called {\it miner}, may receive two types of rewards, comprising {\it subsidy}, the systematically assigned currency, and {\it transaction fees}, attached to the transactions from users. Conventional wisdom has long asserted that subsidy dominates the incentive for mining. However, current Bitcoin has witnessed a prevailing transition from the subsidy-incentive regime to the transaction fee-incentive regime as subsidy of Bitcoin is dwindling (\$50 in 2008 and \$6.25 for now). Since transaction fee is on its way to becoming the material part of the mining reward, a miner's strategy of {\it when} to start up its rig(s) and a user's strategy of {\it how} to bid transaction fee will deviate from the previous belief dramatically. This triggers unprecedented issues that simply do not appear in the subsidy-incentive regime.

\begin{figure}[t]
\centerline{\includegraphics[height=1.6in,width=2.5in]{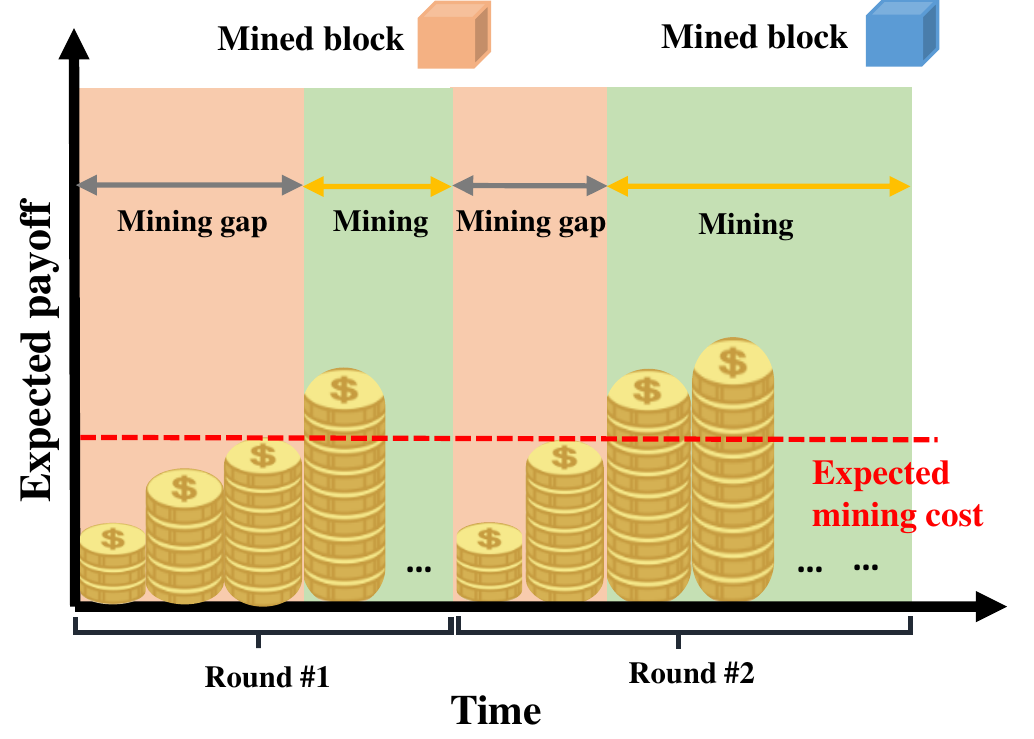}}
\caption{Illustration of mining gaps.  When the expected payoff from mining is lower than the expected cost (the area below the dashed line), rational miners will keep the rigs halted to reduce running costs, waiting for transactions with higher fees, which is denoted as the ``mining gap" period. Only if the expected payoff from mining exceeds the cost, will the mining rigs be activated to work, which is deemed as the ``mining" period. }
\label{gap}
\end{figure}

The subsidy-incentive regime can be viewed as a fixed monetary stimulation mechanism since transaction fee is negligible compared to subsidy. Such a fixed incentive regime makes it profitable for  miners to power on their rigs from the very beginning of each round, since the earlier the miners start up rigs, the more likely they may get the accounting right due to more trials. Nonetheless, being an ``early bird" may be an unfavorable strategy for a miner in the transaction fee-incentive regime. This is because the main source of reward can be adjusted by users in this regime,  a rational miner will stop mining until  transaction fees are sufficient to assure a larger expected mining payoff than the cost. Such a {\it ``no rabbits, no eagles"} strategy incurs the {\it mining gap} \cite{CCS2016,gap}, a period of time in which miners turn the mining rigs off to reduce running cost as shown in Fig. \ref{gap}, rendering no incentives for miners to start up rigs early. Obviously, the ``mining gap" phenomenon 1) aggressively decreases the transaction throughput of Bitcoin, daunting its application in high-concurrency fields and 2) makes the blockchain more vulnerable to attacks, such as double spending attack, forking attack, etc., since the computing power required to execute these attacks is significantly decreased when the honest computing power drops, threatening the security of the blockchain.

However, any user is powerless to shrink the mining gap because the starting up strategy of a miner is not driven by any transaction fee unilaterally, but  the rewards obtained from all transactions in a block. Aware of this, a rational user might provide low transaction fee, being a free-rider. What's worse, in this paper we prove mathematically that 
once there are users who choose to submit low  transaction fees as the optimal strategy, the free-riding phenomenon will be overwhelming in Bitcoin quickly, further expanding the mining gap consequently. We name the troubling case where users are prone to offer low transaction fees and miners are reluctant to start up rigs early as the {\it``egoistic dilemma"}. Undoubtedly, such a predicament ends up weakening the performance and security of blockchain.

\begin{figure}[t]
\centerline{\includegraphics[height=1.3in,width=2.6in]{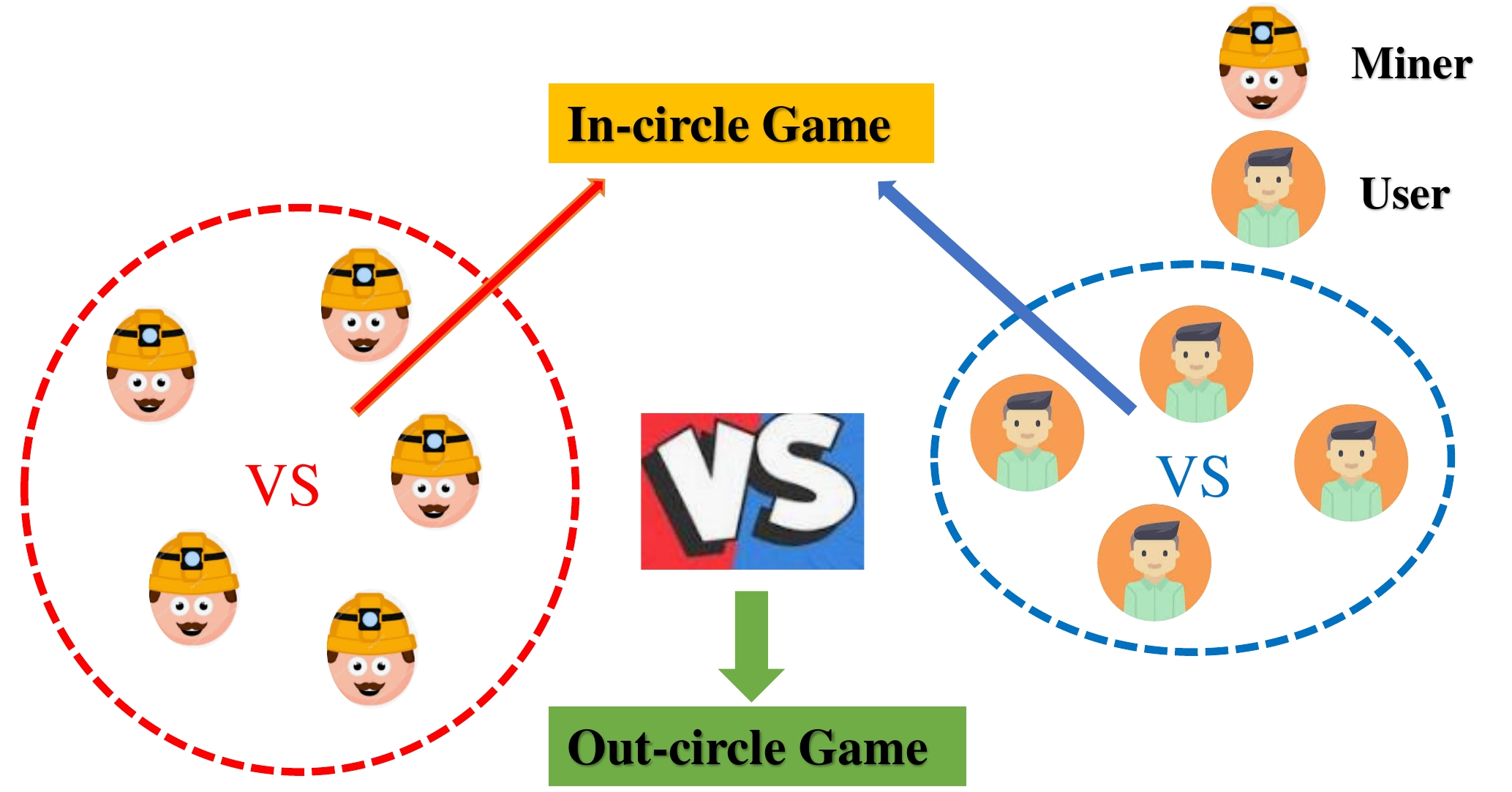}}
\caption{The multi-miner and multi-user game in Bitcoin.}
\label{fig1}
\end{figure}

Until now, we are not aware of any previous recognition to present the egoistic dilemma, which motivates our work to address it in this paper. However, it is challenging to get out of the egoistic dilemma  in that it is the result of a sophisticated multidimensional game. That is to say, confrontational games exist not only between users and miners bilaterally, but also among miners and users internally.  As Fig. \ref{fig1} shows,  we define the  interplay among  homogenous players (i.e., miner-to-miner or user-to-user) as the {\it in-circle game} and the bilateral antagonism between heterogeneous players (i.e., miner-to-user) as the {\it out-circle game}. The nature of multidimensionality (i.e., in-circle and out-circle) makes it hard to coordinate the conflicting interests due to the entangled impacts of each player.

To get rid of the egoistic dilemma in the multi-miner and multi-user game, we first derive the mathematical characteristics of the in-circle game via adopting the supermodular game \cite{superbook}, where the optimal strategies of the profit-driven players are analyzed. This enables us to reasonably untangle the antagonism among homogenous players in the in-circle games,  making all the users (miners) act as a whole, i.e., the user-side (miner-side), to game with their adversary.  Then, we prove that the user-side can utilize the zero-determinant (ZD) strategy \cite{PNAZ} to independently set the expected payoff of the miner-side no matter how it acts. Leveraging this powerful ZD strategy, we devise an incentive mechanism to arm the user-side to induce the miner-side to be an ``early bird", narrowing down the mining gap and extricating the players from the  egoistic dilemma consequently. To the best of our knowledge, we are the first to  present a theoretic study of the mining gap phenomenon and propose an incentive mechanism to solve the egoistic dilemma under the transaction fee-incentive Bitcoin. Conclusively, the contributions of our work can be summarized as follows:

\begin{itemize}

  \item \textbf{Reasonable dimension reduction analysis}.  We prove the interplay among homogenous players  can be described as a supermodular game, based on which the actions of players in the in-circle game is derived to be {\it strategic complementary}. Such a property reveals that rational homogenous players prefer to behave  in unity for  maximizing their profits. Aware of this, we can  reasonably carry out dimension reduction analysis, transforming the complicated multi-miner and multi-user game into a miner-side and user-side game, highly reducing the analysis complexity.
   \item \textbf{Effective ZD-based incentive mechanism}. After simplifying the multidimensional game into a miner-side to user-side one, we propose a ZD-based incentive mechanism which can be laid out to hit the egoistic dilemma hurts. Specifically, the proposed mechanism enables the user-side to lure the miner-side to behave as an ``early bird", disengaging both parties from the egoistic dilemma successfully. Both theoretical analyses and numerical simulations demonstrate the effectiveness of our mechanism.
  \item \textbf{Sustained ability of motivation}. Our ZD-based incentive mechanism  empowers the user-side to drive the miner-side to start up rigs early by increasing the short-term payoffs without any additional payment in the long run.  This guarantees the   user-side to have sustained ability  of monetary  motivating  the miner-side to behave cooperatively.
  \item \textbf{Fairness}.  Even can employ the ZD strategy to  dominate the game, the user-side  has to pay the highest transaction fee for monetary incentivizing the miner-side  to be an  ``early bird", implying that the users in the lead cannot squeeze miners financially. The fairness can make the  miner-side placed at a disadvantage  trust the proposed mechanism,  ensuring its sustainability over the long term.
\end{itemize}

The remainder of the paper is organized as follows. Section \ref{Related Work} lists the related work and Section \ref{Problem statement} formulates the multi-miner and multi-user game. Based on this, the analysis of the in-circle game and that of the out-circle game are respectively conducted in Sections \ref{supermodular game} and \ref{ZD sequential game}. In Section \ref{ZD-based mechanism}, we propose the novel incentive mechanism in light of the ZD theory and testify its effectiveness theoretically. The experimental simulations are carried out in Section \ref{evaluation} and Section \ref{conclusion} concludes our paper finally.

\section{Related Work}\label{Related Work}
To date, transaction fee-incentive Bitcoin has been a largely under-explored domain and research in this field is still in its infancy. In 2016, Carlsten {\it et al.} \cite{CCS2016} first postulated that when transaction fee dominates the mining reward, immediately after a block has been mined, there is zero expected income but nonzero operation cost for the miner \cite{access}. Thus, rational miners would stop mining in this case, resulting in the so-called ``mining gap". Such a phenomenon presents new strategy pattern of the miners, that is, strategically adjust their starting up times of mining for obtaining profitable payoffs, and pioneers new research direction for the transaction fee-incentive Bitcoin.

Inspired by this, Tsabary {\it et al.} \cite{gap} studied miners' strategies in choosing different starting up times by establishing a ``gap game", in which the miner's utility is comprehensively modeled by considering the expected income and cost. Based on this, they concluded that such a gap forms well before fees are the only incentive. Considering that \cite{gap} was carried out under the quasi-static state assumption, Di {\it et al.} \cite{access} analyzed a dynamic game with more realistic hypotheses. Consequently, they mentioned that the decentralization of blockchain is heavily threatened in this case. In addition, \cite{consensus} and \cite{discount} illustrated the above game differently via adopting ``consensus game" and stochastic game, respectively. This provides us another view of investigating the miners' strategic actions as well as the stability of blockchain. Moreover, Arenas {\it et al.} \cite{discount} put forward that the undercutting strategy is a better option for the miners than the honest strategy, i.e., mining on the longest chain, under the transaction fee-incentive regime, which is consistent with the statement in \cite{CCS2016}. While Gong {\it et al.} in \cite{gong} pointed out that contrary to the primary belief, the undercutting behavior is not always beneficial to miners, especially when other miners are aware of these kinds of actions.

In summary, the above studies give the first attempt to analyze the strategy pattern on the transaction fee-incentive regime. However, none of them considers the impact of the users' strategy on miners but only studies the miners' behaviors independently, not to mention the in-depth analysis of our proposed egoistic dilemma. Hence, our work fills the gap in addressing the dilemma of the miners and users, providing the first incentive mechanism to cap the mining gap in the transaction fee-incentive Bitcoin.
\section{Problem Formulation}\label{Problem statement}

In this paper, we consider a Bitcoin system that consists of a set of miners (denoted as $\mathcal{M}$) competing for mining blocks and  a set of users (denoted as $\mathcal{N}$) bidding transaction packaging. That is, a multi-miner and multi-user game is employed to depict the real Bitcoin system. Following the assumption made in \cite{CCS2016,gap}, we suppose each miner has one rig, and all rigs are assumed to be identical which means they share the same operation cost and computing power for simplicity\footnote {Notably, our scheme is also applicable to the case where each miner has multiple or heterogeneous rigs. One can easily extend our model by appropriately adjusting $\alpha_i(t)$ and $C_m(x_i,t)$ defined in the following context.}. Primary parameters used in this section are listed in Table 1.
\subsection{Strategy Model}\label{section3.1}
For any miner $i$, it can choose a starting up time $t_i\in[0,T]$ to launch mining, where $T$ represents the duration from the time when the  previous  block is generated to that when the current  block is created.
We denote the  strategy of miner $i$ as $x_i \in X_i=[0,1]$, in which $x_i=1-\frac{t_i}{T}$. It is obvious that the earlier miner $i$ starts up its rigs, the smaller $t_i$ is and the larger $x_i$ is. Let $y_k \in Y_k=[0,F_h]$ be the strategy of any arbitrary user $k$, which represents the bidding fee it offers, with $F_h$ as the maximum value. In order to analyze the strategies of miners and users, we introduce the following definition.
\begin{definition}[Lattice \cite{superbook}]\label{lattice}
Set $S$ is a lattice if the following three conditions are satisfied:
 \begin{itemize}
  \item S is a partially ordered set.
  \item S has the least upper bound or the greatest lower bound.
  \item $\forall s_1, s_2  \in S$, if we define  $s_1 \cup s_2 =\max(s_1, s_2)$ and $s_1 \cap s_2 =\min(s_1,s_2)$, then the join and meet of each pair should be contained in $S$.
\end{itemize}
\end{definition}

\begin{theorem}\label{theorem 1}
The strategy space $X_i$ of miner $i$ and that of user $k$, i.e., $Y_k$, are lattices.
\end{theorem}
\begin{IEEEproof}
We can prove $X_i$ is a lattice from two steps considering that the second condition is met because of $X_i=[0,1]$. Firstly, $X_i$ should be reflexive, antisymmetric, and transitive in order to be a partially ordered set; secondly, the join and meet of each pair in $X_i$ belong to $X_i$.

According to the definition of $X_i$, all elements are ordered in real line. Hence,  $\forall x_1, x_2, x_3 \in X_i$, we have
$x_1\leq x_1$ (reflexive); if $x_1\leq x_2,x_2\leq x_1$, then $x_1=x_2$ (antisymmetric); if $x_1\leq x_2,x_2\leq x_3$, then $x_1\leq x_2\leq x_3$ (transitive). Thus, $X_i$ is a partially ordered set. Besides, the join and meet of $\forall x_1, x_2 \in X_i$ satisfy $x_1 \cup x_2 =\max(x_1, x_2)\in X_i$ and $x_1 \cap x_2 =\min(x_1,x_2)\in X_i$ \cite{superbook}. Hence, $X_i$ is a lattice. We omit the proof of $Y_k$ for reducing repetition since they are very similar.
\end{IEEEproof}
\begin{table}\label{table1}
\centering
\caption{List of Primary Parameters in Section \ref{Problem statement}.}
\begin{center}
\begin{tabular}{|c|m{6cm}|}
\hline
$\mathcal{M}$, $\mathcal{N}$ & Set of miners and users \\
\hline
$t_i$ & Miner $i$'s starting up time of its rig\\
 \hline
$T$ & The time duration of each round\\
\hline
$B$ & The time when a block is mined\\
\hline
$\lambda$ & Rate parameter of $f_B$\\
\hline
$F_h$ & Maximum fee the user may offer\\
\hline
$x_i$, $y_k$ & Miner $i$ and user $k$ 's strategy\\
\hline
$\boldsymbol{x,y}$ & Vector of strategies of all miners and users\\
\hline
$\boldsymbol{y_{-k}}$ & Vector of strategies of all users expect $k$\\
\hline
$X_i$, $Y_k$ & Miner $i$ and user $k$ 's strategy space\\
\hline
$U_{m_i}$, $U_{u_k}$ & Expected payoff of miner $i$ and user $k$\\
\hline
$\alpha_I(t)$, $\alpha_i(t)$ & Number of active rigs of the system and of miner $i$ at time $t$\\
\hline
$\tau(\cdot)$ & The mining duration of active rigs in the system\\
\hline
$V_m(\cdot), V_u(\cdot)$ & Profit of mining/being packaged for the miner/user\\
\hline
$C_m(\cdot), C_u(\cdot)$ &Cost of mining/bidding for the miner/user\\
\hline
$a,d,g,w$ & The first order partial derivatives of $\alpha_I, \tau, C_m, C_u$\\
\hline
$\varepsilon_m, \varsigma_m, \varepsilon_u,\varsigma_u$ & Positive scaling parameters\\
\hline
\end{tabular}
\end{center}
\end{table}
\subsection{Payoff Model}\label{section3.2}
According to the above analyses, the expected payoff of miner $i$ can be defined as
\begin{equation}\label{gongshi1}
  U_{m_i}=\int_{0}^{\infty} P(profit_i|B=t)\cdot f_B(t,\boldsymbol{x},\lambda)dt.
\end{equation}

In \eqref{gongshi1}, $f_B(t,\boldsymbol{x},\lambda)$ and $P(profit_i|B=t)$ express the probability and profit of miner $i$ when a block is successfully mined at time $B=t$.  According to \cite{gap}, $f_B(t,\boldsymbol{x},\lambda)$ can be formulated as a shift-exponentially distribution. That is,
\begin{equation}\label{gongshi2}
f_B(t,\boldsymbol{x},\lambda)=\lambda \cdot \alpha_I(t)\cdot e^{-\lambda\cdot \tau(\boldsymbol{x},t)},
\end{equation}
where $\alpha_I(t)$ and $\tau(\boldsymbol{x},t)$ present the number of active rigs in the system at time $t$ and the mining duration of them  until time $t$. Besides, $\lambda$ is a rate parameter which is related to the mining difficulty of the system and we denote $\boldsymbol{x}$ as the vector of all the rigs' starting up strategies. Obviously, both $\alpha_I(t)$ and $\tau(\boldsymbol{x},t)$ are monotonically increasing with the starting up strategy, thus we set $\frac{\partial \alpha_I(t)}{\partial x_i}=a>0$ and  $\frac{\partial \tau(\boldsymbol{x},t)}{\partial x_i}=d>0$.

As for miner $i$'s profit, $P(profit_i|B=t)$ can be devised as the difference between the income and expense, namely,
\begin{equation}\label{gongshi3}
P(profit_i|B=t)=\varepsilon_m \frac{\alpha_i(t)}{\alpha_I(t)}\cdot V_m(\boldsymbol{y})-\varsigma_m\cdot C_m(x_i,t),
\end{equation}
where $\alpha_i(t)$ is  the number of active mining rig of miner $i$ at time $t$, thus $\alpha_i(t)\in \{0,1\}$; the ratio between $\alpha_i(t)$ and $\alpha_I(t)$ indicates the success mining rate of miner $i$. Additionally, $V_m(\boldsymbol{y})$ and $C_m(x_i,t)$ show the respective profit and cost of a successful block, in which $\boldsymbol{y}$ denotes all the users' bidding strategies. It is evident that $C_m(x_i,t)$ has a monotonically increasing relationship with regard to $x_i$, thus we set $\frac{\partial C_m(x_i,t)}{\partial x_i}=g>0$. And $\varepsilon_m>0$, $\varsigma_m>0$ are scaling parameters.

With a comparable structure, the expected payoff of user $k$ can be defined as
\begin{equation}\label{gongshi4}
  U_{u_k}=\int_{0}^{\infty} Q(profit_k|B=t) f_B(t,\boldsymbol{x},\lambda) Pro(y_k,\boldsymbol {y_{-k}})dt,
\end{equation}
in which $f_B(t,\boldsymbol{x},\lambda)\cdot Pro(y_k,\boldsymbol{y_{-k}})$ represents the probability that transaction of user $k$ is packaged at time $B=t$ by submitting transaction fee as $y_k$ while others offering $\boldsymbol{y_{-k}}$, and $Q(profit_k|B=t)$ denotes the profit of user $k$. To be specific, $Pro(y_k,\boldsymbol{y_{-k}})$ can be calculated by
$Pro(y_k,\boldsymbol{y_{-k}})=\frac{y_k}{\sum y_{-k}+y_k}$, which satisfies a simple rule that the probability of being packed rises with the increase of $y_k$, but decreases as other users' fees lift. As for user $k$'s profit when its transaction is selected into a mined block, $Q(profit_k|B=t)$ is designed as
\begin{equation}\label{gongshi5}
Q(profit_k|B=t)=\varepsilon_u \cdot V_u(\theta_k)-\varsigma_u\cdot C_u(y_k),
\end{equation}
where $V_u(\theta_k)$ means the income of user $k$ and $\theta_k$ is an endogenous variable showing packaging delay or other impact factors that may influence the user's utility. Besides, $C_u(y_k)$ presents the cost of the user which is monotonically increasing with $y_k$, hence we set $\frac{\partial C_u(y_k)}{\partial y_k}=w>0$. And $\varepsilon_u>0$, $\varsigma_u>0$ are scaling parameters.
\section{Analysis of the In-Circle Game}\label{supermodular game}
As mentioned above, the multi-miner
and multi-user interaction in Bitcoin can be further divided into two kinds of games, i.e., the in-circle and out-circle games.  In this section, we focus on the in-circle game to investigate the interplay of homogenous players. To that aim, we establish a supermodular game-based model to elaborately describe the actions of in-circle players. After that, the mathematical characteristics of players' strategies are figured out, which can be well captured by the term {\it strategic complementarity} \cite{super3}. This allows us to reasonably simplify the multi-miner and multi-user game to a bilateral confrontation game, based on which, an in-depth inspection stressing the interaction between heterogeneous players will be conducted in the next section. In the following, we first introduce the definition of supermodular game.

\begin{definition}[Supermodular Game \cite{superbook}]\label{supergame}
A game  with a set of players $\mathcal{Y}$, strategy space   $(X_v)_{v\in\mathcal{Y}}$  and payoff function  $(U_v)_{v\in\mathcal{Y}}$ (denoted as $\mathcal{G}=[\mathcal{Y},(X_v)_{v\in\mathcal{Y}},(U_v)_{v\in\mathcal{Y}}]$) is a supermodular game if for each player $v\in \mathcal{Y}$:
\begin{itemize}
  \item the strategy space $(X_v)_{v\in\mathcal{Y}}$ is a lattice.
  \item the payoff function $(U_v)_{v\in\mathcal{Y}}$ is continuous in $x_v$ for fixed $\boldsymbol{x_{-v}}$, where $x_v$ and $\boldsymbol{x_{-v}}$ respectively denote the strategy of player $v$  and the strategy vector of other players except $v$.
  \item $(U_v)_{v\in\mathcal{Y}}$ has increasing differences in ($x_v,\boldsymbol{x_{-v}}$), i.e.,$\forall w\ne v,$ $\frac{\partial^2 U_v}{\partial x_v \partial x_w}\ge0$.
\end{itemize}
\end{definition}


\begin{theorem}\label{theorem 2}
The in-circle game among any miner $i\in \mathcal{M}$ and its peers, denoted as $\mathcal{G}_{M_i}=[\mathcal{M},(X_i)_{i\in \mathcal{M}},(U_{m_i})_{i\in \mathcal{M}}]$,  is a supermodular game if   $\alpha_i(t)\cdot\lambda d-a\ge0$ and $g\ge C_m(x_i,t)\cdot\lambda d$ hold.
\end{theorem}
\begin{IEEEproof} We prove this theorem according to Definition \ref{supergame}. 1) Initially, the strategy space of miner $i$, i.e., $X_i$, is a lattice in light of Theorem \ref{theorem 1}; 2) it is clear that $\forall x_i \in X_i$, when $\boldsymbol x_{-i}$ is fixed,  the left and right limits of $U_{m_i}$ are equal, implying that the payoff function of miner $i$ is continuous in $x_i$; 3) we proceed to prove that $\forall x_i, x_j\in X_i, X_j,j\ne i$, $\frac{\partial^2 U_{m_i}}{\partial x_i \partial x_j}\ge 0$ in the following.

Let $P(profit_i|B=t)\cdot f_B(t,\boldsymbol{x},\lambda)=O(t, \boldsymbol{x})$. To prove $\frac{\partial^2 U_{m_i}}{\partial x_i \partial x_j}\ge 0$, we need to testify $\frac{\partial^2 \int_{0}^{\infty} O(t, \boldsymbol{x}) dt}{\partial x_i \partial x_j} = \int_{0}^{\infty}   \frac{\partial^2  O(t, \boldsymbol{x}) }{\partial x_i \partial x_j}dt  \ge 0$ in light of the differentiability property in \cite{integral1,integral2}. Further, if  $\frac{\partial^2 O(t, \boldsymbol{x})}{\partial x_i \partial x_j}\ge0$ holds, the above inequation can be proved by utilizing the integral property in \cite{integral2}. Hence, we are going to find the conditions to meet $\frac{\partial^2 O(t, \boldsymbol{x})}{\partial x_i \partial x_j}\ge0$.
In detail, the explicit mathematical form of $O(t, \boldsymbol{x})$ can be written as:
\begin{equation}\label{gongshi6}
\begin{aligned}
  &O(t, \boldsymbol{x})=[\varepsilon_m \frac{\alpha_i(t)}{\alpha_I(t)} V_m(\boldsymbol y)-\varsigma_m C_m(x_i,t)][\lambda \alpha_I(t) e^{-\lambda \tau(\boldsymbol{x},t)}]\\
  &=[\varepsilon_m\alpha_i(t) V_m(\boldsymbol y)\lambda e^{-\lambda \tau(\boldsymbol{x},t)}]\\
  &+[-\varsigma_m C_m(x_i,t)\alpha_I(t) \lambda e^{-\lambda \tau(\boldsymbol{x},t)}].
\end{aligned}
\end{equation}

Next, we divide \eqref{gongshi6} into two parts for convenience of expression, recognized as $\Delta_1+\Delta_2$. Then, the second order partial derivatives of $\Delta_1$ and $\Delta_2$ are respectively \begin{equation}\label{gongshi7}
\begin{aligned}
\frac{\partial^2 \Delta_1}{\partial x_i \partial x_j}&=\varepsilon_m\cdot a \cdot V_m(\boldsymbol y)\cdot\lambda e^{-\lambda\tau(\boldsymbol{x},t)}\cdot[-\lambda d]\\
&+\varepsilon_m\cdot\alpha_i(t)\cdot V_m(\boldsymbol y)\cdot \lambda e^{-\lambda\tau(\boldsymbol{x},t)}\cdot\lambda d\cdot\lambda d\\
&=\varepsilon_m\cdot V_m(\boldsymbol y)\cdot \lambda e^{-\lambda\tau(\boldsymbol{x},t)}\cdot\lambda d\cdot[\alpha_i(t)\cdot\lambda d-a],
\end{aligned}
\end{equation}


\begin{equation}\label{gongshi9}
\begin{aligned}
\frac{\partial^2 \Delta_2}{\partial x_i \partial x_j}=&\varsigma_m\cdot g\cdot \lambda e^{-\lambda\tau(\boldsymbol{x},t)}\cdot [\alpha_I(t)\cdot\lambda d-a]\\
+&\varsigma_m \cdot C_m(x_i,t) \cdot \lambda e^{-\lambda\tau(\boldsymbol{x},t)}\lambda d\cdot[2a-\alpha_I(t)\cdot\lambda d].
\end{aligned}
\end{equation}

Thus, if
$\alpha_i(t)\cdot\lambda d-a\ge0$ and $g\ge C_m(x_i,t)\cdot\lambda d$, we have $\frac{\partial^2 (\Delta_1+\Delta_2)}{\partial x_i \partial x_j}\ge0$. Therefore, Theorem \ref{theorem 2} is proved.

\end{IEEEproof}


\begin{theorem}\label{theorem 3}
The in-circle game among any user $k\in \mathcal{N}$ and its peers, denoted as  $\mathcal{G}_{U_k}=[\mathcal{N},(Y_k)_{k\in \mathcal{N}},(U_{u_k})_{k\in \mathcal{N}}]$,  is a supermodular game when $[\varepsilon_u\cdot V_u(\theta_k)-\varsigma_u\cdot C_u(y_k)]\cdot(y_k-\sum y_{-k})+(y_k+\sum y_{-k})y_k\varsigma_u w\ge0$ is satisfied.
\end{theorem}

\begin{IEEEproof}
The proof is similar to that of Theorem \ref{theorem 2}. It can be easily obtained that $\mathcal{G}_{U_k}$ satisfies the first two conditions of Definition \ref{supergame} to be a supermodular game. Thus, we focus on the third one  i.e., $\forall y_k, y_l \in Y_k, Y_l, l\ne k, \frac{\partial^2 U_{u_k}}{\partial y_k \partial y_l}\ge 0$. Set $Q(profit_k|B=t)\cdot f_B(t,\boldsymbol{x},\lambda)\cdot Pro(y_k,\boldsymbol {y_{-k}}) = \Delta_3$, we have
\begin{equation}\label{gongshi10}
\begin{aligned}
\frac{\partial^2 \Delta_3}{\partial y_k \partial y_l}&=[\varepsilon_u\cdot V_u(\theta_k)-\varsigma_u\cdot C_u(y_k)]\\
&\cdot\frac{(y_k+\sum y_{-k})^2-2(y_k+\sum y_{-k})\sum y_{-k}}{(y_k+\sum y_{-k})^4}\\
&+\frac{y_{k}\varsigma_u w}{(y_k+\sum y_{-k})^2}\\
&=[\varepsilon_u\cdot V_u(\theta_k)-\varsigma_u\cdot C_u(y_k)]\cdot \frac{y_k-\sum y_{-k}}{(y_k+\sum y_{-k})^3}\\
&+\frac{y_k\varsigma_u w}{(y_k+\sum y_{-k})^2}.
\end{aligned}
\end{equation}

Hence, when $[\varepsilon_u\cdot V_u(\theta_k)-\varsigma_u\cdot C_u(y_k)]\cdot(y_k-\sum y_{-k})+(y_k+\sum y_{-k})y_k\varsigma_u w\ge0$, the theorem holds.
\end{IEEEproof}

According to \cite{superbook,super2,super3}, supermodular game $\mathcal{G}_{M_i}, i\in \mathcal{M} $ ($\mathcal{G}_{U_k}, k\in \mathcal{N} $) indicates that the marginal payoff of any miner $i$ (user $k$) can be increased if all players in the in-circle game behave in unity. That is to say, when other miners select late starting up strategies, it becomes more profitable for miner $i$ to power on its rigs lately and vice versa; similarly, if other users bid low transaction fees, offering a low transaction fee as well will benefit user $k$ and vice versa. Such a complementary strategy occurs in homogenous players is termed as {\it strategic complementarity} \cite{super3} and we can use another metaphor, i.e., {\it theatre effect}, to illustrate it more intuitively: when most of the audience stand up to watch a movie, the best response of the other is to stand up as well, keeping the same pace with others.

\section{Analysis of the Out-Circle Game}\label{ZD sequential game}
The strategic complementarity of the in-circle game presents that a rational player should  act synchronously with others for maximizing its payoff. As a result, once there are users who choose to provide low transaction fees and become free-riders, all other users will select the same strategy, enlarging the mining gap consequently, as mentioned in Section \ref{Intro}. However, the strategic complementarity is a double-edged sword since it can serve to reshape the out-circle game from the multi-miner and multi-user game to the miner-side and user-side one, in which all miners share one strategy  $x\in X=[0,1]$ and all users own one strategy $y\in Y=[0,F_h]$ due to the synchronous actions among homogenous players. In such a game, the payoff functions of the miner-side and user-side can be devised as:
\begin{equation}\label{gongshi12}
S_{m}(x,y)=\varpi_m\cdot \chi_m(y)-\varkappa_m\cdot \Xi_m(x),
\end{equation}
\begin{equation}\label{gongshi11}
S_{u}(x,y)=\varpi_u\cdot \chi_u(x)-\varkappa_u\cdot \Xi_u(y).
\end{equation}

Specifically, $\chi_m(y)$ and $\chi_u(x)$ are profit functions for the miner-side and user-side, which are monotonically increasing with the user-side's bidding devotion $y$ and the miner-side's starting up strategy $x$. Besides, $\Xi_m(x)$ and $\Xi_u(y)$ are the cost functions, sharing the increasing relationship with variables $x$ and $y$, respectively. And $\varpi_m,\varkappa_m,\varpi_u,\varkappa_u$ are positive scaling parameters.
Additionally, compared with \eqref{gongshi1} and \eqref{gongshi4}, we omit the probabilities of mining blocks successfully of the miner-side and being packaged of the user-side, i.e., $f_B(t,\boldsymbol{x},\lambda)$ and $f_B(t,\boldsymbol{x},\lambda)\cdot Pro(y_k,\boldsymbol {y_{-k}})$ in \eqref{gongshi12} and \eqref{gongshi11}. This is because no matter which miner mines a block and which user's transaction is selected, it means that the miner-side mines successfully and the transactions of the user-side are packaged. That is to say, $f_B(t,\boldsymbol{x},\lambda)=1$ and $f_B(t,\boldsymbol{x},\lambda)\cdot Pro(y_k,\mathbf {y_{-k}})=1$ in this case.
The parameters used in this section is summarized in Table II.

\begin{table}\label{table2}
\centering
\caption{List of Primary Parameters in Section \ref{ZD sequential game}.}
\begin{center}
\begin{tabular}{|c|m{5.5cm}|}
\hline
$x,y$ & Strategy of the miner-side and user-side\\
\hline
$X,Y$ & Strategy space of the two sides\\
\hline
$S_m(\cdot)$, $S_u(\cdot)$ & Payoff function of the two sides\\
\hline
$\chi_m(\cdot), \chi_u(\cdot) $ & Profit function of the two sides\\
\hline
$\Xi_m(\cdot), \Xi_u(\cdot)$ & Cost function of the two sides\\
\hline
$\varpi_m, \varkappa_m, \varpi_u,\varkappa_u$ & Positive scaling parameters\\
\hline
$q^{\pi}(y|X_{-1},Y_{-1})$ & Mixed strategy of the user-side in round $\pi$\\
\hline
$p^{\pi}(x|y)$ & Mixed strategy of the miner-side in round $\pi$\\
\hline
$\mathbf{\Gamma}^{\pi}$& Markov transition matrix in round $\pi$\\
\hline
$z^{\pi}$ & Joint transition probability in round $\pi$\\
\hline
$E_m^{\pi}, E_u^{\pi}$ & Expected payoff of the two sides in round $\pi$\\
\hline
$\eta_1, \eta_2$ & The number of partitions of $X$ and $Y$\\
\hline
$\mu_1, \mu_2$ & The partition interval of $X$ and $Y$\\
\hline
$\mathbf{S_M}, \mathbf{S_U}$ & Payoff vector of the two sides in the discrete form\\
\hline
$q^{\pi}_{ab-s}, p^{\pi}_{s-r}$ & Mixed strategy of the user-side and miner-side in discrete form\\
\hline
\end{tabular}
\end{center}
\end{table}

In the out-circle game, the user-side sets its transaction fee as the bidding strategy first. After observing current transaction fee, the miner-side chooses a profitable starting up strategy subsequently. Note that the game is led by the user-side and then followed by the miner-side, it is therefore called a {\it sequential game.}
\begin{theorem}\label{theorem 4}
An egoistic dilemma exists in the sequential game between the user-side  and miner-side.
\end{theorem}

\begin{IEEEproof}
To begin with, $\frac{\partial S_{u}}{\partial y}=-\varkappa_u\cdot \frac{\partial \Xi_u(y)}{\partial y}<0$ holds because both $\varkappa_u$ and $\frac{\partial \Xi_u(y)}{\partial y}$ are positive. This means the  payoff of the user-side is inversely proportional to its strategy $y$, thus offering the minimum fee no matter what strategy the miner-side selects is the  best strategy of the user-side. Comparatively, we also obtain $\frac{\partial S_{m}}{\partial x}=-\varkappa_m\cdot \frac{\partial \Xi_m(x)}{\partial x}<0$ because of $\varkappa_m>0$ and $\frac{\partial \Xi_m(x)}{\partial x}>0$. Such a reverse relationship between the miner-side's payoff and its strategy results in the best response of the miner-side being $0$. Hence, the equilibrium of the game is $(x^*,y^*)=(0,0)$, which is an unexpected outcome for the players compared with the state $(x,y)=(1,F_h)$. Accordingly, an egoistic dilemma exists in the sequential game between the user-side  and miner-side.
\end{IEEEproof}

Aware of the above theorem, we can state that the user-side and miner-side may be trapped in the egoistic dilemma, where the user-side is not willing to submit high transaction fee and the miner-side is reluctant to start up rigs early. In the short run, such egoistic strategies decrease social welfare for both parties, and meanwhile, it may lead to low efficiency or even break down the Bitcoin system in the long run. To address this problem, we equip the user-side with a novel incentive mechanism to lure the miner-side's early starting up strategy. The reason why we facilitate the user-side instead of the miner-side is that in the sequential game, it is the user-side that acts first, showing its advantages in influencing the miner-side's action. Hence, by teaching the miner-side to get the cognition that starting up earlier may bring about higher payoffs, the user-side is capable of incentivizing the interest-driven miner-side to wake up rigs early. Such an aim can be greatly achieved by leveraging the revolutionary sequential ZD strategy presented in \cite{ZDhu,PNAZ,MTD}.

The classical ZD strategy \cite{PNAZ} offers valuable insights to understand the Markov games, which enables the ZD adopter can unilaterally set the expected payoff of its adversary no matter how the adversary acts. Such a capability in controlling the expected payoff of others allows us to enact an incentive mechanism to induce the miner-side to be an ``early bird". To that aim, we need to answer two fundamental questions:
\begin{itemize}
\item can the user-side be a ZD adopter?
\item what should the user-side do to incentivize the miner-side to start up rigs at the earliest time?
\end{itemize}

To answer the first question, we employ the Markov game to model the interaction between the user-side and miner-side. We first assume that the user-side makes decisions based on the outcomes of previous rounds, while the miner-side sets its strategy after knowing the user-side's strategy in this round. In light of this, we define the mixed strategy of the user-side in round $\pi$ as $q^\pi(y|X_{-1},Y_{-1})$, which denotes the probability density\footnote{The reason why we use probability density rather than probability is that the strategy spaces of our scenario are continuous.} of offering $y$ as the bidding fee when the previous outcome is $X_{-1}Y_{-1}$ and $\int_0^{F_h} q^\pi(y|X_{-1},Y_{-1})dy=1$ holds. Comparably, the mixed strategy of the miner-side is thereby described as $p^\pi(x|y)$, expressing the probability density of starting up rigs with strategy $x$ after realizing the user-side's fee is provided as $y$ in round $\pi$ and also, we have $\int_{0}^{1} p^\pi(x|y)dx=1$. Hence, the Markov matrix $\mathbf{\Gamma}^{\pi}$ representing the state transition probability density from round $\pi-1$ to $\pi$ can be deduced as
$\mathbf{\Gamma}^{\pi}(X_{-1},Y_{-1},x,y)=q^\pi(y|X_{-1},Y_{-1})\cdot p^\pi(x|y)$. Accordingly, the joint probability density that the user-side adopting strategy $y$ while the miner-side choosing strategy $x$ in round $\pi$ can be denoted as $z^\pi(x,y)=z^{\pi-1}(X_{-1},Y_{-1})\cdot \mathbf{\Gamma}^{\pi}(X_{-1},Y_{-1},x,y)$. Therefore, the expected payoffs of the user-side and miner-side in round $\pi$ can be presented as:
\begin{equation}\label{gongshi13}
E_{u}^\pi=\int_{0}^{F_h}\int_{0}^{1}z^\pi(x,y)\cdot S_{u}(x,y)dxdy,
\end{equation}
\begin{equation}\label{gongshi14}
E_{m}^\pi=\int_{0}^{F_h}\int_{0}^{1}z^\pi(x,y)\cdot S_{m}(x,y)dxdy.
\end{equation}

Subsequently, we divide the strategy spaces of the miner-side and user-side as $\{0,\mu_1,...,\eta_1\mu_1\}$ and $\{0,\mu_2,...,\eta_2\mu_2\}$ with sufficiently small $\mu_1, \mu_2$ and large enough $\eta_1,\eta_2$, satisfying $\eta_1\mu_1=1$ and $\eta_2\mu_2=F_h$. Thus, the strategy spaces can be approximated to continuous ones if $\mu_1, \mu_2 \rightarrow 0$ and $\eta_1,\eta_2 \rightarrow \infty$. Accordingly, the payoffs of the two players can be partitioned as $\mathbf{S_M}=[S_{m}(0,0),...,S_{m}(\eta_1\mu_1,0),...,
S_{m}(\eta_1\mu_1,\eta_2\mu_2)]$ and
$\mathbf{S_U}=[S_{u}(0,0),...,S_{u}(\eta_1\mu_1,0),...,
S_{u}(\eta_1\mu_1,\eta_2\mu_2)]$. Based on this, we can rewrite the mixed strategy of the user-side in the discrete form  as $q_{ab-s}^\pi$ for round $\pi$, in which $a\in\{0,1,...,\eta_1\},b,s\in\{0,1,...,\eta_2\}$, representing the probability of submitting $s\mu_2$ as the bidding fee in round $\pi$  when the user-side provides the payment as $b\mu_2$ while the miner-side choosing strategy $a\mu_1$ in the previous round. Similarly, the mixed strategy of the miner-side can be thereby depicted as $p_{s-r}^\pi$ in round $\pi$, where $s\in\{0,1,...,\eta_2\},r\in\{0,1,...,\eta_1\}$, denoting the probability of choosing strategy $r\mu_1$ currently with knowing $s\mu_2$ as the user-side's strategy.

Based on the partitioned strategy spaces and payoffs above, we can conclude that the user-side can control the expected payoff of the miner-side through being a ZD adopter, which is demonstrated in the following theorem.

\begin{theorem}\label{theorem 5}
When the user-side sets its strategy as $q^\pi(F_h|X_{-1},Y_{-1})=\alpha \mathbf{S_{M}}+\beta \mathbf{S_{U}}+\gamma\mathbf{1}$, the expected payoffs of the user-side and miner-side meet a linear relationship, i.e., $\alpha E_{m}^\pi+\beta E_{u}^\pi+\gamma=0$, in which $\alpha, \beta, \gamma$ are coefficients and $q^\pi(F_h|X_{-1},Y_{-1})$ satisfies \begin{equation}
\label{gongshi16}
q^\pi(F_h|X_{-1},Y_{-1})=\\
\left\{
\begin{aligned}
q^\pi(F_h|X_{-1},Y_{-1})&,&X_{-1}<1, \\
q^\pi(F_h|X_{-1},Y_{-1})-1&,&X_{-1}=1.
\end{aligned}
\right.
\end{equation}
\end{theorem}

\begin{IEEEproof}
Firstly, the Markov matrix $\mathbf{\Gamma}^{\pi}$ can be explicitly expressed as
$\mathbf{\Gamma}^{\pi}=[\mathbf{\Gamma}^{\pi}_{00},...,\mathbf{\Gamma^{\pi}_{\eta_1\eta_2}}]$, where $\mathbf{\Gamma}^{\pi}_{rs},r\in\{0,1,...,\eta_1\},s\in\{0,1,...,\eta_2\}$ represents the vector consisting of the transition possibilities from all the previous states $\{X_{-1},Y_{-1}\}$ to current state $\{X=r\mu_1, Y=s\mu_2\}$, i.e.,
$\mathbf{\Gamma}^{\pi}_{rs}=[q_{00-s}^\pi p_{s-r}^\pi,...,q_{\eta_1\eta_2-s}^\pi p_{s-r}^\pi]$.
According to the ZD strategy \cite{PNAZ}, if the stationary vector of $\mathbf{\Gamma}^{\pi}$ is $\boldsymbol{\sigma}$, then we can obtain $\boldsymbol{\sigma^{T}}\cdot\mathbf{\Gamma}^{\pi}=\boldsymbol{\sigma^{T}}$ and $\boldsymbol{\sigma^{T}}\cdot\mathbf{\Gamma^{\pi}}'=\mathbf{0}$
where $\mathbf{\Gamma^{\pi}}'=\mathbf{\Gamma}^{\pi}-\mathbf{I}$ with $\mathbf{I}$ as the unitary matrix. In light of Cramer's law, we get $Adj(\mathbf{\Gamma^{\pi}}')\mathbf{\Gamma^{\pi}}'=det(\mathbf{\Gamma^{\pi}}'\mathbf{I})=0$, where $Adj(\mathbf{\Gamma^{\pi}}')$ is the adjugate matrix of $\mathbf{\Gamma^{\pi}}'$. Hence, we can define an arbitrary $\eta_1\eta_2$-vector, i.e., $\boldsymbol{\zeta}=[\zeta_{00},...,\zeta_{\eta_1\eta_2}]^T$, and conduct dot product of the stable vector $\boldsymbol{\sigma}$ with $\boldsymbol{\zeta}$ to obtain a new Markov matrix after elementary column transformation \cite{PNAZ,ZDhu}, which is
\begin{equation}\label{gongshi15}
\boldsymbol{\sigma}\cdot\boldsymbol{\zeta}=\left[
  \begin{matrix}
   q_{00-0}^{\pi}p_{00}^{\pi}&\cdots&q_{00-\eta_2}^{\pi}&\zeta_{00}\\
   \vdots&\vdots&\vdots&\vdots&\\
   q_{\eta_10-0}^{\pi}p_{00}^{\pi}&\cdots&q_{\eta_10-\eta_2}^{\pi}-1&\zeta_{\eta_10}\\
   \vdots&\vdots&\vdots&\vdots&\\
   q_{\eta_1\eta_2-0}^{\pi}p_{00}^{\pi}&\cdots&q_{\eta_1\eta_2-\eta_2}^{\pi}-1&\zeta_{\eta_1\eta_2}
  \end{matrix}
  \right].
\end{equation}

From \eqref{gongshi15}, one can easily get that the penultimate column is solely determined by the user-side, which can be denoted as $\boldsymbol{\hat{q^\pi}}=[q_{00-\eta_2}^\pi,...,
q_{\eta_10-\eta_2}^\pi-1,...,q_{\eta_1\eta_2-\eta_2}^\pi-1]^T$.
Thus, when the user-side sets its strategy, i.e., $\boldsymbol{\hat{q^\pi}}$, equal to the last column, we can obtain $\boldsymbol{\sigma}\cdot\boldsymbol{\zeta}=0$ since there are two identical columns in a determinant. Further, when $\boldsymbol{\zeta}=\alpha\mathbf{S_M}+\beta\mathbf{S_U}+\gamma\mathbf{1}$, $\boldsymbol{\sigma}\cdot\boldsymbol{\zeta}=\boldsymbol{\sigma}\cdot(\alpha\mathbf{S_M}+\beta\mathbf{S_U}+\gamma\mathbf{1})
=\alpha E_{m}^\pi+\beta E_{u}^\pi+\gamma$ holds \cite{PNAZ}. Thus, the user-side is capable of setting strategy $\boldsymbol{\hat{q^\pi}}$ equal to $\alpha\mathbf{S_M}+\beta\mathbf{S_U}+\gamma\mathbf{1}$ and resulting in a linear relationship between the expected payoffs of the user-side and miner-side, that is, $\alpha E_{m}^\pi+\beta E_{u}^\pi+\gamma=0$. Therefore, we obtain Theorem \ref{theorem 5}.
\end{IEEEproof}

Now, we can answer the first question by stating that the user-side is capable of being a ZD adopter by appropriately setting its strategy according to Theorem \ref{theorem 5}, so as to unilaterally determine the expected payoff of the miner-side as $E_{m}^\pi=-\frac{\gamma}{\alpha}\in[\min E_{m}^\pi,\max E_{m}^\pi]$ when $\beta$ is set as 0. It is also worth to note that in sequential games, only the leader can employ the powerful ZD strategy to control the outcome of the game while the follower is not adequate to do so \cite{ZDhu}.
\section{ZD-based Incentive Mechanism}\label{ZD-based mechanism}
In this section, we are going to answer the second question by equipping the user-side with a ZD-based incentive mechanism. To begin with, we consider the miner-side to be brainy whose strategy is adaptive and iterative rather than fixed and permanent. That is to say, the rational miner-side may learn the utilities of different strategies and adjust its actions towards the best strategy which leads to favorable payoffs while keeping away from the ones that trigger disadvantageous payoffs. Such a property in strategy selection is rooted in the players since they are born with the nature of ``seeking profit and avoiding harm", which is similar to the principle of ``survival of the fittest" in biological evolution. Hence, we consider an {\it evolutionary} miner-side in this paper, who can reasonably adjust its strategy to maximize its payoff by evaluating the corresponding payoffs under different strategies.

We give an example of the evolutionary strategy inspired by \cite{evolution1,evolution2} and claim that other evolutionary strategies shown in \cite{not} and so on share the same mathematical core. Let the probability that the miner-side starts up rigs at the earliest time in round $\pi$ be
$\tilde{p}_{e}^\pi$, then such a probability in round $\pi+1$ iterates as follows:
\begin{equation}\label{gongshi19}
\begin{aligned}
\tilde{p}_{e}^{\pi+1}=\tilde{p}_{e}^\pi\cdot \frac{W_{e}^\pi}{E_{m}^\pi}.
\end{aligned}
\end{equation}

In \eqref{gongshi19},  $W_{e}^\pi$ indicates the expected payoff of the miner-side in round $\pi$ when its strategy is $x=1$, which can be deduced by $W_{e}^\pi=\sum_{s=0}^{\eta_2}g_{s\mu_2}^{\pi}S_m(1,s\mu_2)$, 
and $g_{s\mu_2}^{\pi}$ represents the frequency of the user-side choosing $s\mu_2, s\in\{0,1,...,\eta_2\}$ as its strategy in round $\pi$. Besides, $E_{m}^\pi$ presents the total expected payoff of the miner-side in round $\pi$, calculated by 
$E_{m}^\pi=\sum_{r=0}^{\eta_1}f_{r\mu_1}^{\pi}W_{r\mu_1}^{\pi}$, and $f_{r\mu_1}^{\pi}$ is estimated by the frequency of the miner-side setting $r\mu_1, r\in \{0,1,...,\eta_1\}$ as its strategy in round $\pi$. In addition, $W_{r\mu_1}^{\pi}$ is the expected payoff when the miner-side chooses $r\mu_1$ as its strategy, and is defined as $W_{r\mu_1}^{\pi}=\sum_{s=0}^{\eta_2}g_{s\mu_2}^{\pi}\cdot S_m(r\mu_1,s\mu_2)$. 

\begin{algorithm}[t]\label{suanfa}
\caption{ZD-based Incentive Mechanism}
\begin{algorithmic}[1]\REQUIRE ~~\\ 
The state transition probability matrix of the miner-side, which is deduced according to the preliminary $R$ rounds, $\{P_{a \rightarrow r}\}_{\eta_1\times \eta_1}$;\\
The number of iterations $Q$;\\
The maximum and minimum payoffs of the miner-side controlled by the user-side, $\max E_{m}^\pi$ and $\min E_{m}^\pi$;\\
\STATE Set $E_m^0=\frac{1}{\eta_1+1}\Sigma_{a=0}^{\eta_1}P_{a \to \eta_1}*(\max E_{m}^\pi-\min E_{m}^\pi)+\min E_{m}^\pi$ and $\kappa=\frac{1}{\eta_1+1}\Sigma_{a=0}^{\eta_1}P_{a \to \eta_1}$
\FOR{$\iota=1$ to $Q$}
\IF{$x^{\iota-1}=\rho \mu_1,\forall \rho \in \{0,1,...,\eta_1\} $}\label{line2}
  \IF{$P_{\rho \rightarrow \eta_1}\ge P_{\rho\rightarrow\theta},  \forall\theta\in\{0,1,...,\eta_1-1\}$}\label{line3}
    \STATE $\kappa=(1+P_{\rho \rightarrow \eta_1})\cdot \kappa$
    \STATE Calculate $q^{\iota}$ to set $E_{m}^{\iota}=\max E_{m}^\pi \cdot \frac{1}{1+e^{-\omega_1 \cdot \kappa}}$

  \ELSE
  \STATE $\kappa=(1+1/P_{\rho \rightarrow \eta_1})\cdot \kappa$
    \STATE Calculate $q^{\iota}$ to set $E_{m}^{\iota}=\min E_{m}^\pi \cdot (\frac{1}{1+e^{\omega_2 \cdot \kappa}}+1)$
  \ENDIF
\ENDIF
\IF {the present round ends}
 \STATE Update $\{P_{a\rightarrow r}\}_{\eta_1\times \eta_1}$
\ENDIF
\ENDFOR
\end{algorithmic}
\end{algorithm}

In the following, we devise a ZD-based incentive mechanism for the user-side to drive the miner-side to be an ``early bird", which is illustrated in Algorithm 1. Conclusively, the essence of our mechanism is to reward the miner-side with a higher payoff when it starts up rigs at the earliest time, yet penalizing it by setting its payoff as a lower one if it begins to mine lately. Leveraging such a market regulation mechanism, the user-side can make the miner-side perceive the positive relationship between the earlier starting up strategy and the more profitable utility, encouraging the miner-side to be the earliest strategy adopter eventually.

We elaborate the pseudo-code in Algorithm 1 detailedly as follows. Considering that the user-side acts first to set the miner-side's payoff without seeing its strategy, it is necessary for the user-side to predict the possible strategy of its opponent in each round. Hence, we set a preliminary phase containing $R$ rounds to estimate the state transition probability matrix $\{P_{a\rightarrow r}\}_{\eta_1\times \eta_1}$ as the basis of predicting the miner-side's action, in which $P_{a\rightarrow r}, a,r \in\{0,1,...,\eta_1\}$ denotes the probability that the miner-side transits from previous strategy $a\mu_1$ to current strategy $r\mu_1$. To be specific, $P_{a\rightarrow r}$ in round $\pi+1$ can be calculated by
$P_{a\rightarrow r}=P^{\pi+1}(r|a)=\frac{p^{\pi+1}(r)\cdot p^{\pi}(a|r)}{p^{\pi+1}(a)}$, where $p^{\pi+1}(r)$ and $p^{\pi+1}(a)$ respectively refer to the probability of choosing strategy $r\mu_1$ and $a\mu_1$ in round $\pi+1$.

After obtaining the prediction of $\{P_{a\rightarrow r}\}_{\eta_1\times \eta_1}$, the user-side will set the miner-side's payoff for each round accordingly. To be concrete, the user-side will initially reward the miner-side proportionally to its estimated average probability of being an ``early bird", i.e., $\frac{1}{\eta_1+1}\Sigma_{a=0}^{\eta_1}P_{a \to \eta_1}$ and set the parameter $\kappa$ (Line 1). After that, when the previous strategy of the miner-side is regarded as $\rho\mu_1, \forall \rho \in \{0,1,...,\eta_1\}$ (Line \ref{line2}), then whether the miner-side is an ``early bird" or not can be deduced from the following two cases: if the transition probability from the previous strategy to the earliest strategy $P_{\rho \rightarrow \eta_1}$ is no less than that of any other possible transitions, i.e., $P_{\rho \rightarrow \eta_1}\ge P_{\rho\rightarrow\theta}, \forall\theta\in\{0,1,...,\eta_1-1\}$ (Line 4), then the miner-side can be deemed as an ``early bird", who will be rewarded with a higher payoff by setting $E_m^{\iota}=\max E_{m}^\pi \cdot \frac{1}{1+e^{-\omega_1 \cdot \kappa}}$,  where $\kappa=(1+P_{\rho \rightarrow \eta_1})\cdot \kappa$ and $\omega_1>0$ is a scaling parameter (Line 5-6). Note that the higher $P_{\rho \rightarrow \eta_1}$ is, the more increment of $\kappa$ will have, and the higher reward the miner-side could possess, until the maximum value of the miner-side could obtain. On the other hand, if $\exists \theta \in \{0,1,...,\eta_1-1\}$, such that $P_{\rho \rightarrow \eta_1}<P_{\rho \rightarrow \theta}$ (Line 7), the user-side in this case may utilize the ZD strategy to penalize the miner-side for giving a lower payoff by making $E_{m}^{\iota}=\min E_{m}^\pi \cdot (\frac{1}{1+e^{\omega_2 \cdot \kappa}}+1)$, in which $\kappa=(1+1/P_{\rho \rightarrow \eta_1})\cdot \kappa$ and $\omega_2>0$ is a scaling parameter (Line 8-9). It is worth to note that the lower $P_{\rho \rightarrow \eta_1}$ is, the higher $\kappa$ is and the lower payoff the miner-side is rewarded, until the minimum value it could possess. When each round ends, the user-side will collect the game results and recalculate $\{P_{a \rightarrow r}\}_{\eta_1\times \eta_1}$ for a more precise estimation of the miner-side's next action (Line 12-14). The above procedures repeat until $Q$ rounds are carried out.


\begin{theorem}[Effectiveness]\label{theorem 6}
For any evolutionary miner-side who is encouraged by the ZD-based incentive mechanism, its probability of choosing the earliest starting up strategy becomes to 1 at last, i.e., $\tilde{p}_{e}^\pi \to 1$.
\end{theorem}

\begin{IEEEproof}
According to \eqref{gongshi19}, $\tilde{p}_{e}^{\pi+1}$ increases only when $W_{e}^\pi>E_{m}^{\pi}$. Thus, we proceed to prove $W_{e}^\pi>E_{m}^{\pi}$ from two cases, where the miner-side is recognized as an ``early bird" or not.

Case 1): if the miner-side is regarded as an ``early bird" in round $\pi+1$ with strategy $x^{\pi+1}=1$, then the user-side will set $E_{m}^{\pi+1}> E_{m}^{\pi}$. Under this case, the miner-side's probability of choosing the earliest strategy and the corresponding expected payoff come to
\begin{equation*}
\begin{aligned}
f_{e}^{\pi+1}=\frac{f_{e}^{\pi}(R+\pi)+1}{R+\pi+1}
=f_{e}^{\pi}+\frac{1-f_{e}^{\pi}}{R+\pi+1},
\end{aligned}
\end{equation*}
and
\begin{equation*}
\begin{aligned}
E_{m}^{\pi+1}=\sum_{r=0}^{\eta_1-1}f_{r\mu_1}^{\pi+1}\cdot W_{r\mu_1}^{\pi+1}+f_{e}^{\pi+1}\cdot W_{e}^{\pi+1}.
\end{aligned}
\end{equation*}

Since the probability of a miner-side's strategy $f_{r\mu_1}^{\pi+1}, r\in\{0,1,...,\eta_1-1\}$ is not higher than the sum of possibilities of all the non-earliest actions $1-f_{e}^{\pi+1}$,  $f_{r\mu_1}^{\pi+1}\leq1-f_{e}^{\pi+1}$ holds. Therefore, we have
\begin{equation*}\label{W}
\begin{aligned}
W_{e}^{\pi+1}&=\frac{E_{m}^{\pi+1}-\sum_{r=0}^{\eta_1-1}f_{r\mu_1}^{\pi+1}\cdot W_{r\mu_1}^{\pi+1}}{f_{e}^{\pi+1}}\\
&=\frac{E_{m}^{\pi+1}-\sum_{r=0}^{\eta_1-1}f_{r\mu_1}^{\pi+1}\cdot W_{r\mu_1}^{\pi+1}}{f_{e}^{\pi}+\frac{1-f_{e}^{\pi}}{R+\pi+1}}.
\end{aligned}
\end{equation*}

When $R+\pi\to\infty$, because of $E_{m}^{\pi+1}> E_{m}^{\pi}$ and $W_{r\mu_1}^{\pi+1}=W_{r\mu_1}^\pi,\forall r\in\{0,1,...,\eta_1-1\}$, $W_{e}^{\pi+1}$ turns to
\begin{equation}\label{W2}
\begin{aligned}
W_{e}^{\pi+1}
>\frac{E_{m}^{\pi}-\sum_{r=0}^{\eta_1-1}f_{r\mu_1}^{\pi}\cdot W_{r\mu_1}^{\pi}}{f_{e}^{\pi}}=W_{e}^\pi.
\end{aligned}
\end{equation}

Case 2): when the miner-side's strategy in round $\pi+1$ is predicted as  $x^{\pi+1}=r^*\mu_1,r^*\in\{0,1,...,\eta_1-1\}$, the miner-side's expected payoff is thereby controlled as $E_m^{\pi+1}<E_m^{\pi}$. Accordingly, the miner-side's probability of selecting to be an ``early bird" or not are respectively

\begin{equation*}
\begin{aligned}
f_{e}^{\pi+1}=\frac{f_{e}^{\pi}(R+\pi)}{R+\pi+1}
=f_{e}^{\pi}-\frac{f_{e}^{\pi}}{R+\pi+1},
\end{aligned}
\end{equation*}
and
\begin{equation*}
\begin{aligned}
f_{r^*\mu_1}^{\pi+1}=\frac{f_{r^*\mu_1}^{\pi}(R+\pi)+1}{R+\pi+1}
=f_{r^*\mu_1}^{\pi}+\frac{1-f_{r^*\mu_1}^{\pi}}{R+\pi+1},
\end{aligned}
\end{equation*}
with $ f_{r\mu_1}^{\pi+1}=f_{r\mu_1}^{\pi}, \forall r\ne r^*, r\in\{0,1,...,\eta_1-1\}.$

Hence, we can get
\begin{equation*}
\begin{aligned}
&\sum_{r=0}^{\eta_1-1}f_{r\mu_1}^{\pi+1}\cdot W_{r\mu_1}^{\pi+1}=E_{m}^{\pi+1}-f_{e}^{\pi+1}\cdot W_{e}^{\pi+1}\\
&=E_{m}^{\pi+1}-(f_{e}^{\pi}-\frac{f_{e}^{\pi}}{R+\pi+1})\cdot W_{e}^{\pi+1}.
\end{aligned}
\end{equation*}

when $R+\pi\to\infty$, on account of $W_{e}^{\pi+1}=W_{e}^\pi$, we obtain
\begin{equation*}
\begin{aligned}
\sum_{r=0}^{\eta_1-1}f_{r\mu_1}^{\pi+1}\cdot W_{r\mu_1}^{\pi+1} = E_{m}^{\pi+1}-f_{e}^{\pi} W_{e}^{\pi+1}<\sum_{r=0}^{\eta_1-1}f_{r\mu_1}^{\pi} W_{r\mu_1}^{\pi}.
\end{aligned}
\end{equation*}

Since $\forall r \ne r^*, r\in\{0,1,...,\eta_1-1\}$, $f_{r\mu_1}^{\pi+1}$ remains unchanged and $f_{r^*\mu_1}^{\pi+1}$ rises. Besides, we have
\begin{equation}\label{3}
\begin{aligned}
\sum_{r=0}^{\eta_1-1}f_{r\mu_1}^{\pi+1}\cdot W_{r\mu_1}^{\pi+1}
&=\sum_{r=0}^{r^*-1}f_{r\mu_1}^{\pi+1}\cdot W_{r\mu_1}^{\pi+1}\\
&+ f_{r^*\mu_1}^{\pi+1}\cdot W_{r^*\mu_1}^{\pi+1}
+\sum_{r=r^*+1}^{\eta_1-1}f_{r\mu_1}^{\pi+1}\cdot W_{r\mu_1}^{\pi+1}\\
&<\sum_{r=0}^{\eta_1-1}f_{r\mu_1}^{\pi} W_{r\mu_1}^{\pi}.
\end{aligned}
\end{equation}

Hence, we have $W_{r^*\mu_1}^{\pi+1}<W_{r^*\mu_1}^{\pi}$ and $W_{r\mu_1}^{\pi+1}<W_{r\mu_1}^{\pi},\forall r\ne r^*, r\in\{0,1,...,\eta_1-1\}$. Consequently, we have $W_{r\mu_1}^{\pi+1}<W_{r\mu_1}^\pi, \forall r\in\{0,1,...,\eta_1-1\}$.

To sum up, Case 1) indicates $W_{e}^{\pi}$ raises while $W_{r\mu_1}^{\pi}, r\in\{0,1,...,\eta_1-1\}$ holds unchanged as the round goes up; Case 2) suggests that $W_{r\mu_1}^{\pi}, r\in\{0,1,...,\eta_1-1\}$ decreases with $W_{e}^{\pi}$ keeping stable. Hence, $\exists \pi>\pi^*$, such that $W_{e}^{\pi}>W_{r\mu_1}^{\pi}, r\in\{0,1,...,\eta_1-1\}$. Based on which, $E_{m}^{\pi}$ turns to
\begin{equation*}\label{4}
\begin{aligned}
E_{m}^{\pi}&=\sum_{r=0}^{\eta_1-1}f_{r\mu_1}^{\pi}\cdot W_{r\mu_1}^{\pi}+f_{e}^{\pi}\cdot W_{e}^{\pi}\\
&<\sum_{r=0}^{\eta_1-1}f_{r\mu_1}^{\pi}\cdot W_{e}^{\pi}+f_{e}^{\pi}\cdot W_{e}^{\pi}=W_{e}^\pi.
\end{aligned}
\end{equation*}

Hence, when $E_{m}^\pi<W_{e}^\pi$ is satisfied, $\tilde{p}_{e}^{\pi+1}$ equals to $\tilde{p}_{e}^\pi\cdot \frac{W_{e}^\pi}{E_{m}^\pi}\to 1$, resulting in the inevitability of the miner-side being an ``early bird". Therefore, we obtain Theorem \ref{theorem 6}.
\end{IEEEproof}

In light of Theorem \ref{theorem 6}, one may raise a question that {\it if the user-side is so powerful to guide the miner-side to be an ``early bird", will it be possible to squeeze the miner-side financially by offering low transaction fee to achieve greedy purpose?} We address the above issue by presenting the following theorem.

\begin{theorem}[Fairness]\label{theorem 7}
When the miner-side is driven to choose the earliest starting up strategy, the only rational strategy of the user-side who employs the ZD strategy is to provide the highest transaction fee.
\end{theorem}
\begin{IEEEproof}
According to Theorem \ref{theorem 5} and \cite{PNAZ}, we have

\begin{equation}\label{gongshi21}
\left\{
\begin{aligned}
q_{00-\eta_2}^{\pi} &= \alpha \cdot E_{m}^{\pi}(0,0)+\gamma,\\
q_{\eta_1\eta_2-\eta_2}^{\pi}-1&=\alpha \cdot E_{m}^{\pi}(1,F_h)+\gamma. \\
\end{aligned}
\right.
\end{equation}

We can use $q_{00-\eta_2}^{\pi}$ and $q_{\eta_1\eta_2-\eta_2}^{\pi}$ to describe $\alpha$ and $\gamma$ as
\begin{equation}\label{gongshi21}
\left\{
\begin{aligned}
\alpha&=\frac{q_{00-\eta_2}^{\pi}
-q_{\eta_1\eta_2-\eta_2}^{\pi}+1}
{E_{m}^{\pi}(0,0)-E_{m}^{\pi}(1,F_h)},\\ \gamma&=\frac{(q_{\eta_1\eta_2-\eta_2}^{\pi}-1)
E_{m}^{\pi}(0,0)-q_{00-\eta_2}^{\pi}
E_{m}^{\pi}(1,F_h)}{E_{m}^{\pi}(0,0)-E_{m}^{\pi}(1,F_h)}.\\
\end{aligned}
\right.
\end{equation}

Hence, the range of $E_{m}^\pi$ can be derived as $E_{m}^\pi=-\frac{\gamma}{\alpha}=\frac{
(1-q_{\eta_1\eta_2-\eta_2}^{\pi})E_{m}^{\pi}(0,0)
+q_{00-\eta_2}^{\pi}E_{m}^{\pi}(1,F_h)}
{1-q_{\eta_1\eta_2-\eta_2}^{\pi}+q_{00-\eta_2}^{\pi}}$ \cite{PNAZ, ZDhu}.  Accordingly, when  $q_{00-\eta_2}^{\pi}=1$ and $q_{\eta_1\eta_2-\eta_2}^{\pi}=1$, $E_{m}^\pi$ gets the maximum, i.e., $\max E_{m}^\pi$= $E_{m}^{\pi}(1,F_h)$. The upper bound $E_{m}^{\pi}(1,F_h)$ demonstrates that when the miner-side is lured to select the earliest starting up strategy, the user-side will set its strategy as $y=F_h$ to reward the miner-side with the maximum expected payoff. 
Thus, the proposed ZD-based incentive mechanism makes no room for the user-side to behave greedily through financially squeezing the miner-side, showing its fairness to both sides.
\end{IEEEproof}

\begin{theorem}[Sustained ability of motivation]\label{theorem 8}
In the long run, the actual payoff of the miner-side is equivalent to $E_{m}^{\pi}(1,F_h)$.
\end{theorem}
\begin{IEEEproof}
Based on Theorem \ref{theorem 6} and Theorem \ref{theorem 7}, we can conclude that $\exists \pi^+\in \mathbb{Z^+}$ such that $\pi>\pi^+$, the probability that the miner-side starts up rigs at the earliest time is 1, i.e., $\tilde{p}_{e}^\pi=1, \forall \pi>\pi^+$. In this case, the expected payoff of the miner-side is $E_{m}^{\pi}(1,F_h)$. Accordingly, the actual payoff of the miner-side $\Upsilon_m$ can be derived as the average of the expected payoffs $E_{m}^{t}$ in which $t<\pi^+$ and $E_{m}^{t}=E_{m}^{\pi}(1,F_h)$ where $t\ge\pi^+$. Hence, $\Upsilon_m$ can be calculated as
\begin{equation}\label{gongshi23}
\begin{aligned}
\Upsilon_m=\lim\limits_{\pi\to\infty}\frac{\Sigma_{t=1}^{\pi^+-1}E_{m}^{t}+\Sigma_{t=\pi^+}^{\pi} E_{m}^{\pi}(1,F_h)}{\pi}=E_{m}^{\pi}(1,F_h).
\end{aligned}
\end{equation}

\end{IEEEproof}

Notably, Theorem \ref{theorem 7} and Theorem \ref{theorem 8} conclusively indicate that the maximum payoff of the miner-side in each round which is controlled by the user-side, i.e, $\max{E_{m}^{\pi}}$ equals to the actual payoff of the miner-side over the long run, i.e., $E_{m}^{\pi}(1,F_h)$. This reveals that only through offering transaction fee generously as $F_h$ in each round, is the user-side capable of incentivizing the miner-side to be an ``early bird" in the long run without any additional payment. Moreover, $F_h$ does not exceed the maximum payment the user-side can afford, which ensures the user-side to have sustained ability to monetarily incentivize the miner-side to behave collaboratively.


\section{Experimental Evaluations}\label{evaluation}
\begin{figure}[t]
\centering
\subfigure[$\mathbf{p^0}=\mathbf{0.3},\mathbf{q^0}=\mathbf{0.5}$.]{
\begin{minipage}[t]{0.5\linewidth}
\centering
\includegraphics[height=1.3in,width=1.73in]{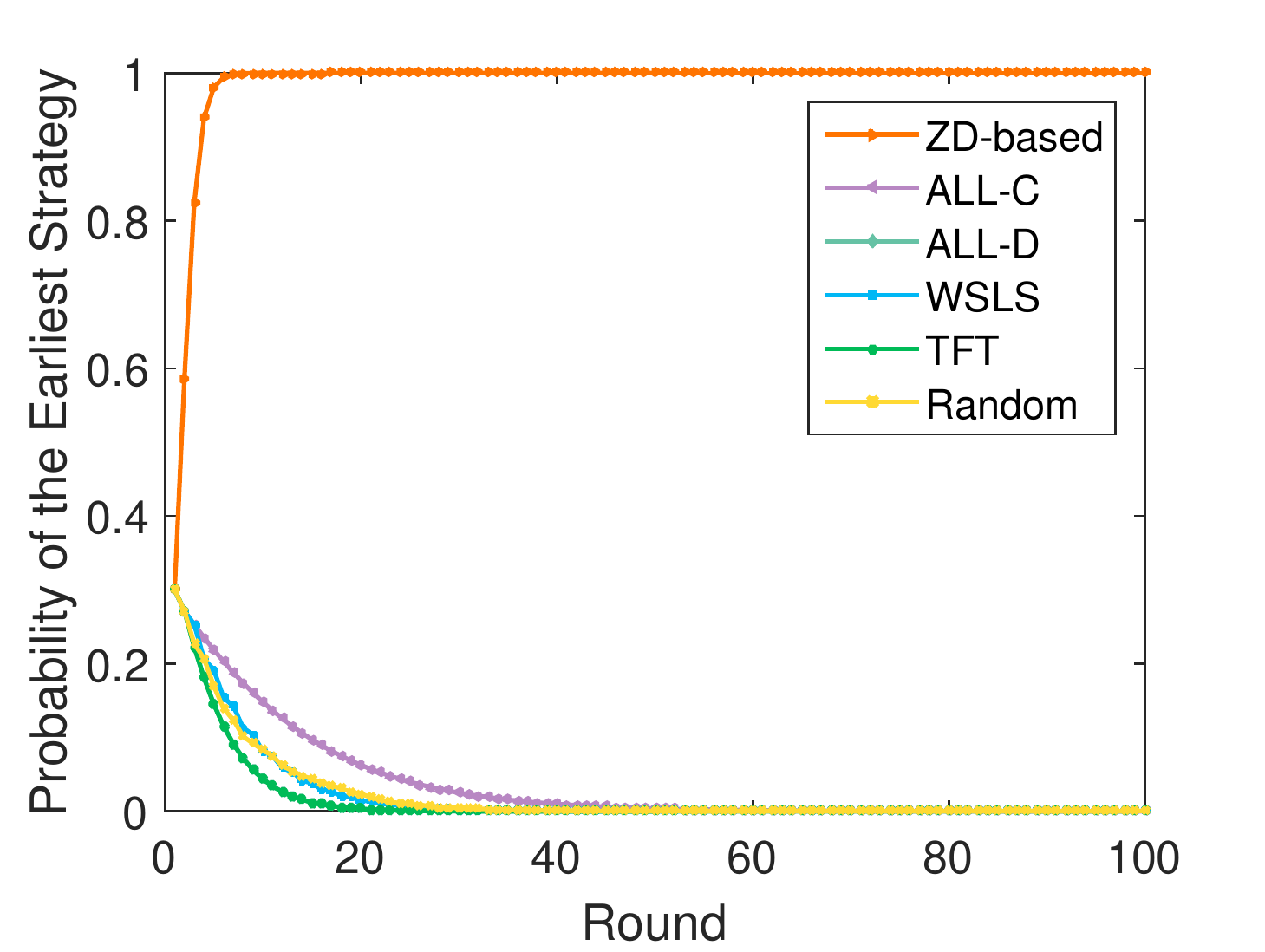}
\end{minipage}%
}%
\subfigure[$\mathbf{p^0}=\mathbf{0.5},\mathbf{q^0}=\mathbf{0.7}$.]{
\begin{minipage}[t]{0.5\linewidth}
\centering
\includegraphics[height=1.3in,width=1.73in]{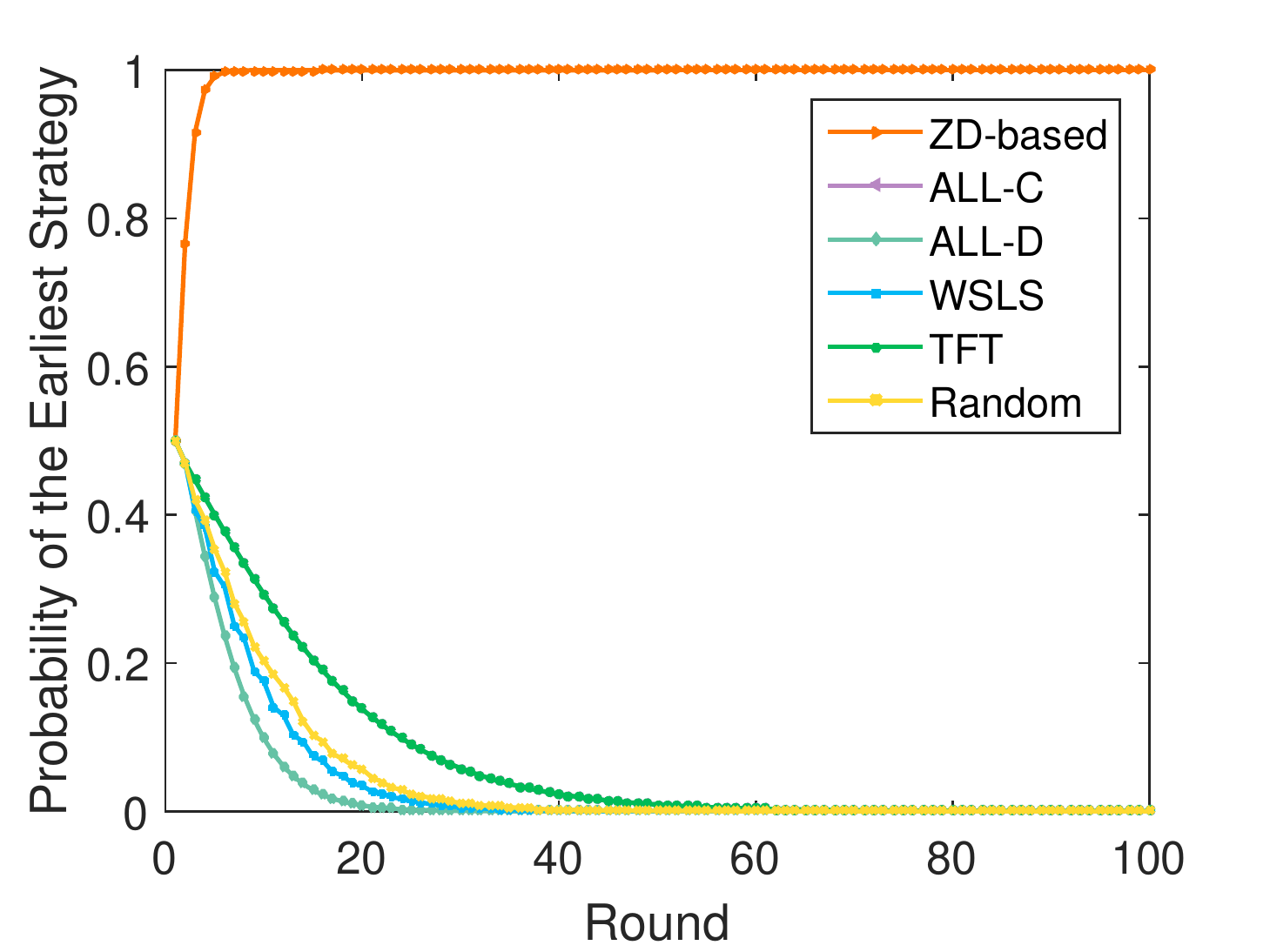}
\end{minipage}%
}%

\subfigure[$\mathbf{p^0}=\mathbf{0.7},\mathbf{q^0}=\mathbf{0.9}$.]{
\begin{minipage}[t]{0.5\linewidth}
\centering
\includegraphics[height=1.3in,width=1.73in]{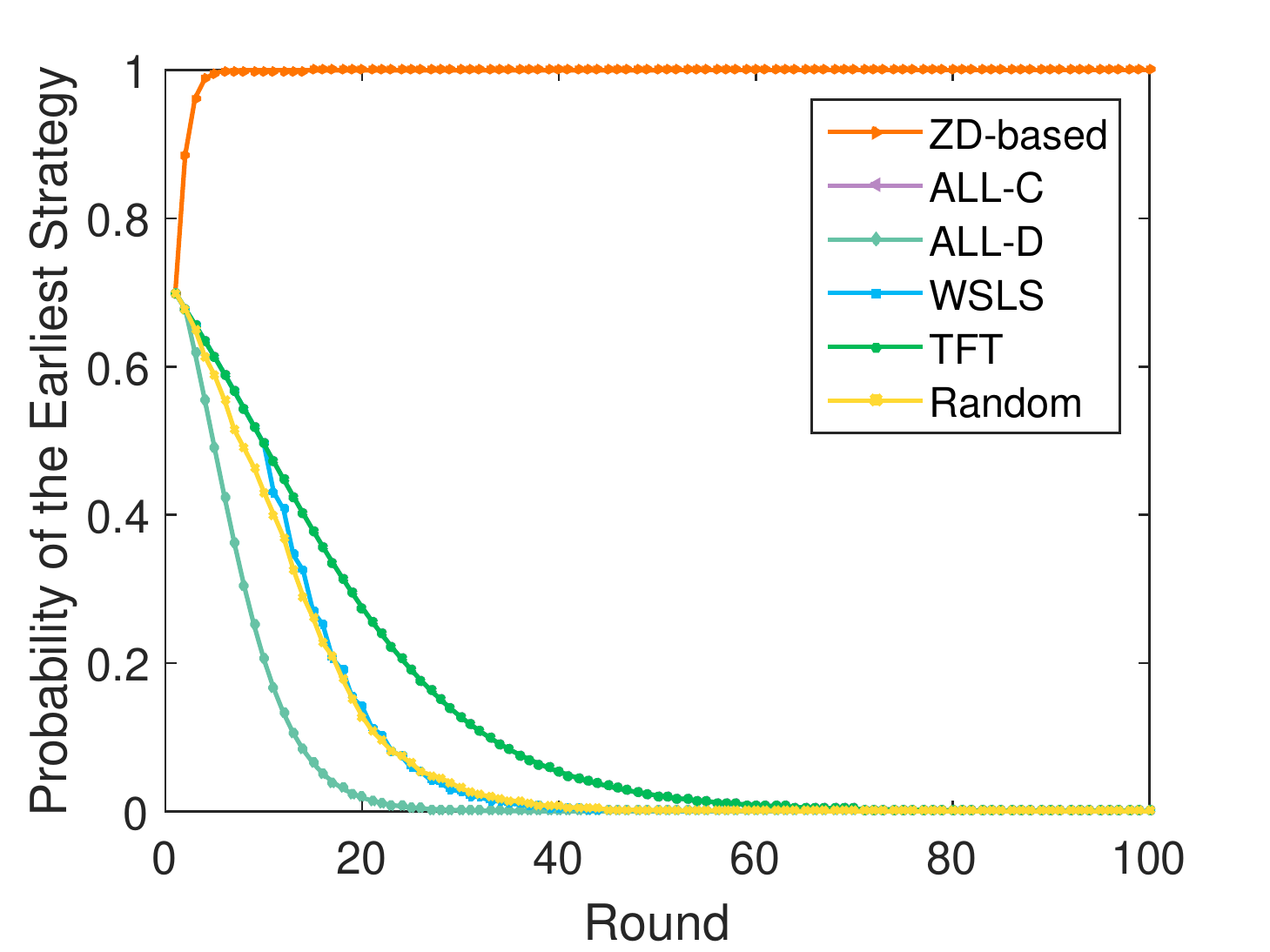}
\end{minipage}
}%
\subfigure[$\mathbf{p^0}=\mathbf{0.9},\mathbf{q^0}=\mathbf{0.3}$.]{
\begin{minipage}[t]{0.5\linewidth}
\centering
\includegraphics[height=1.3in,width=1.73in]{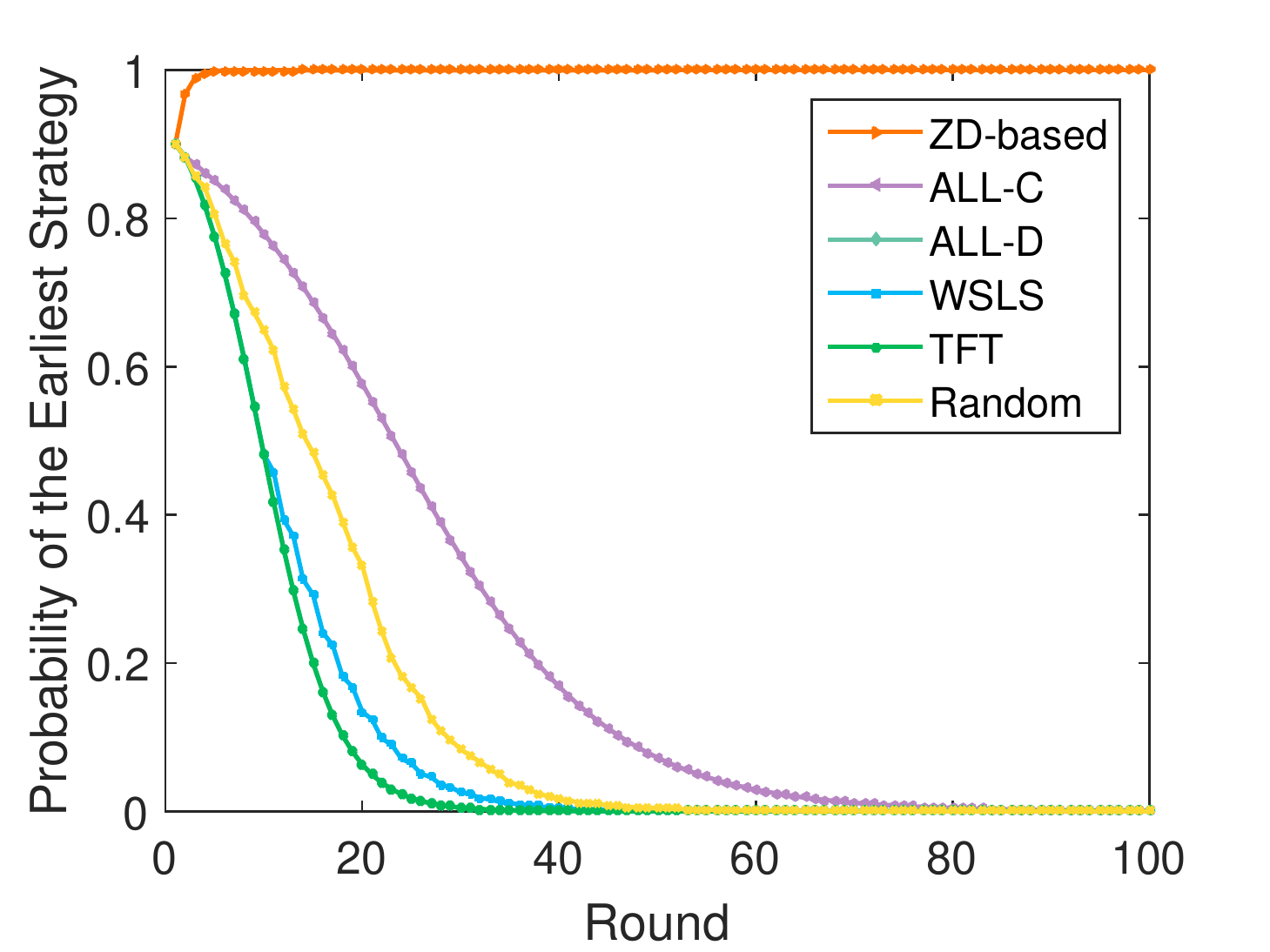}
\end{minipage}
}%
\centering
\caption{The evolutions of the earliest starting up probabilities of the miner-side when the user-side adopts different strategies.}
\label{fig2}
\end{figure}

\begin{figure}[t]
\centering
\subfigure[Probability of choosing the earliest starting up strategy of the miner-side.]{
\begin{minipage}[t]{0.48\linewidth}
\centering
\includegraphics[height=1.3in,width=1.73in]{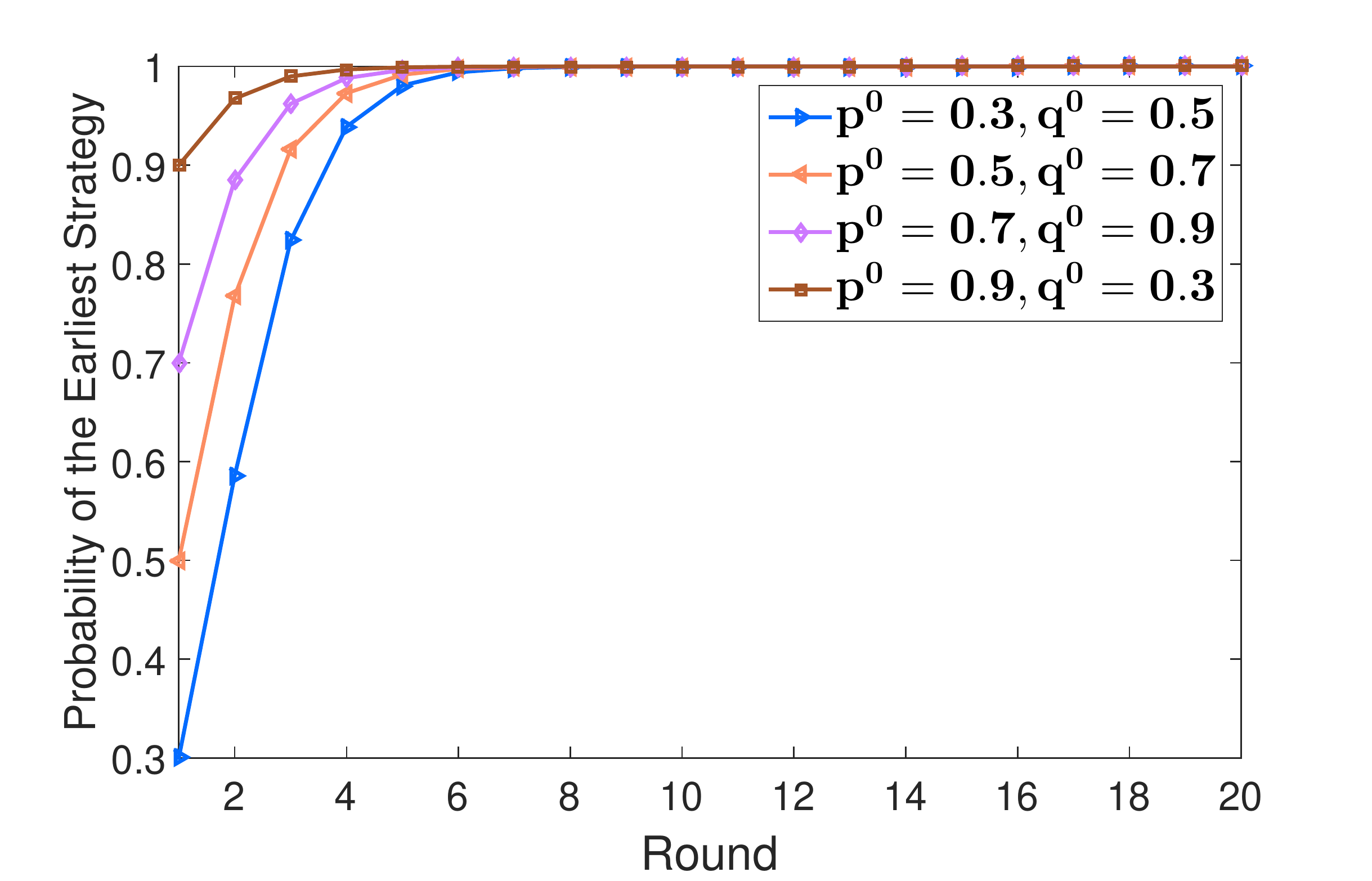}
\end{minipage}%
}%
\subfigure[Starting up time of the miner-side.]{
\begin{minipage}[t]{0.48\linewidth}
\centering
\includegraphics[height=1.3in,width=1.73in]{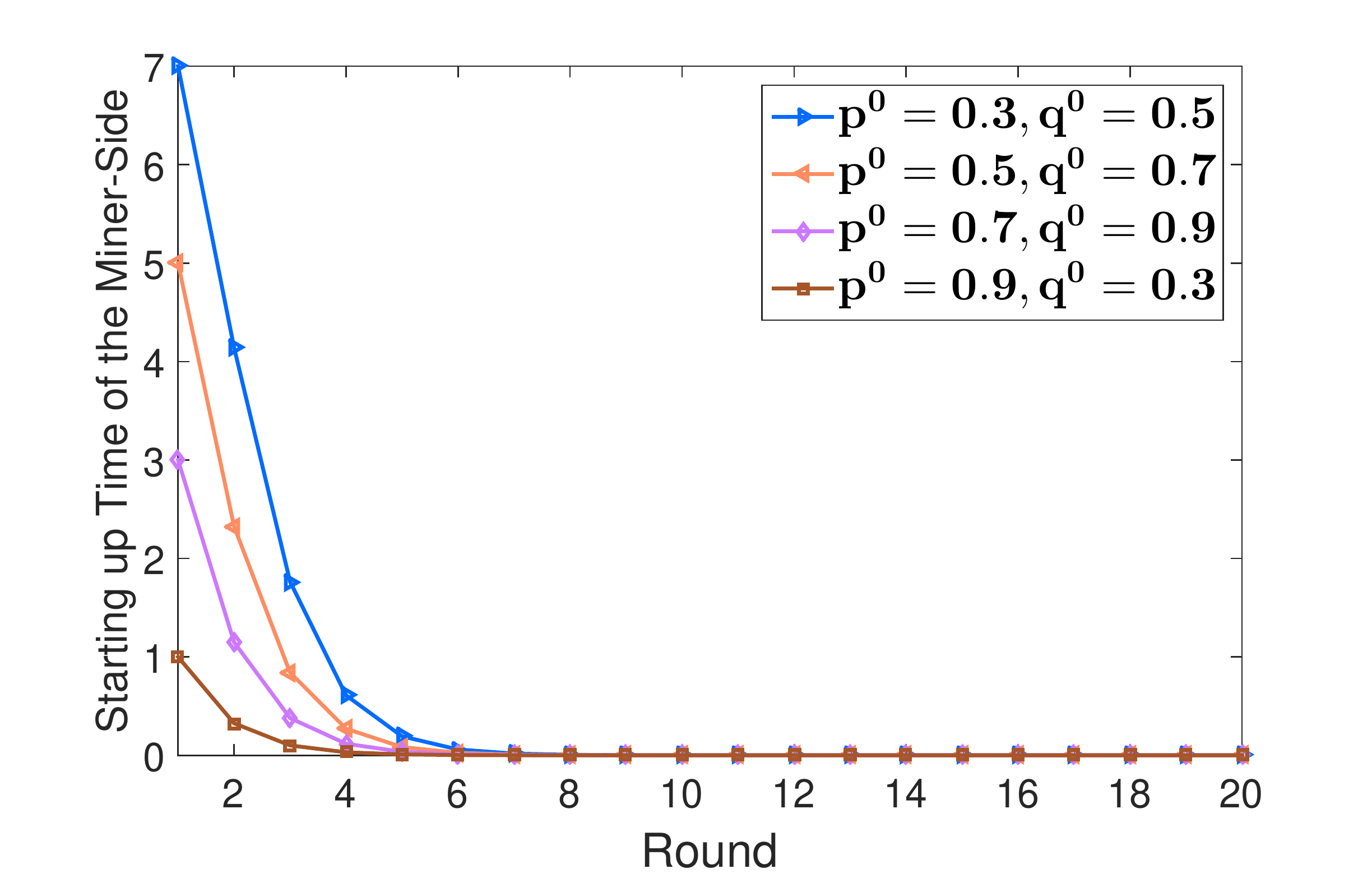}
\end{minipage}%
}%
\centering
\caption{The evolutions of the miner-side vs. various initial probability settings.}
\label{fig4}
\end{figure}

\begin{figure}[t]
\centering
\subfigure[$\mathbf{p^0}=\mathbf{0.3},\mathbf{q^0}=\mathbf{0.5}$.]{
\begin{minipage}[t]{0.48\linewidth}
\centering
\includegraphics[height=1.3in,width=1.73in]{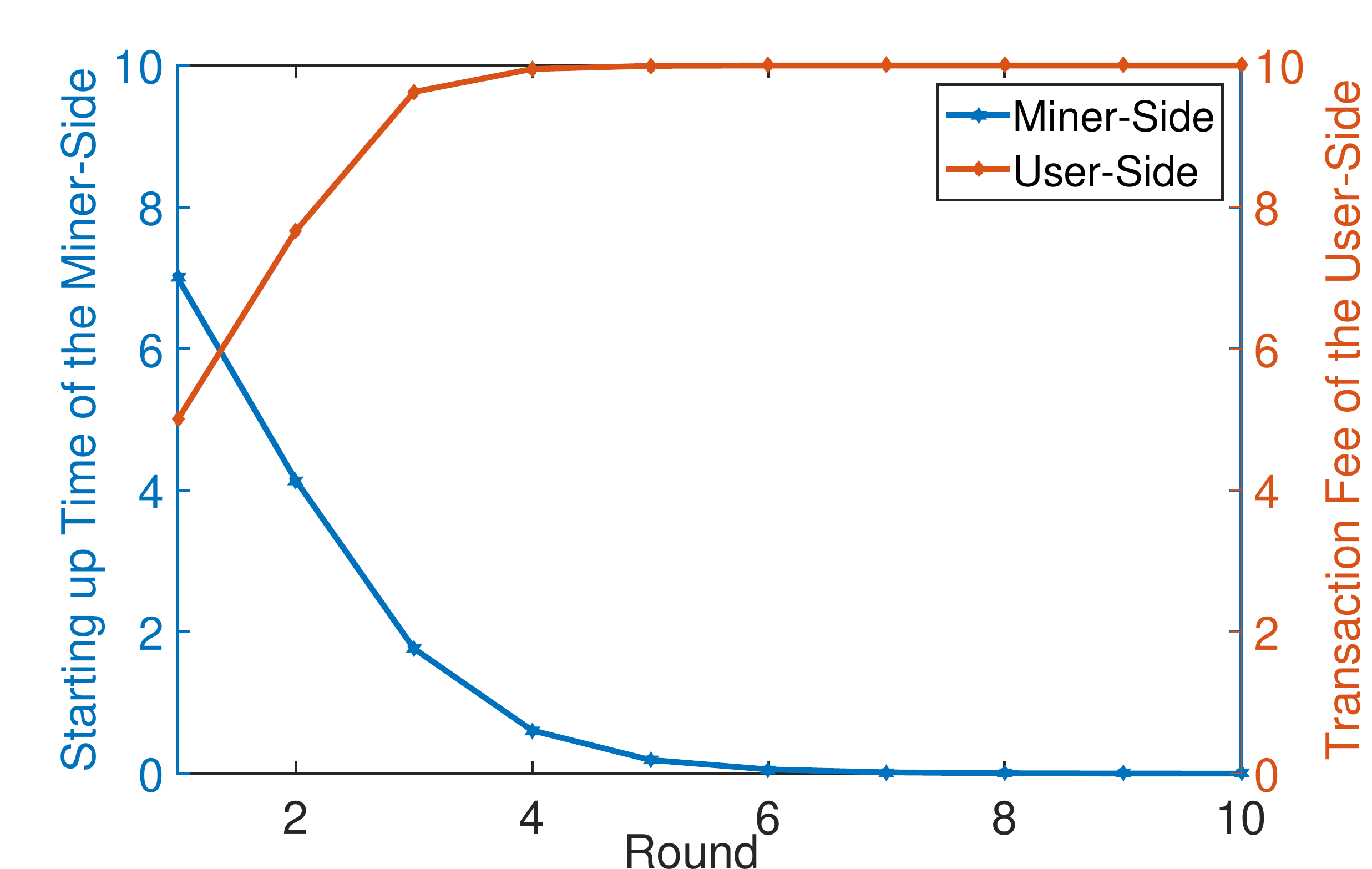}
\end{minipage}%
}%
\subfigure[$\mathbf{p^0}=\mathbf{0.5},\mathbf{q^0}=\mathbf{0.7}$.]{
\begin{minipage}[t]{0.48\linewidth}
\centering
\includegraphics[height=1.3in,width=1.73in]{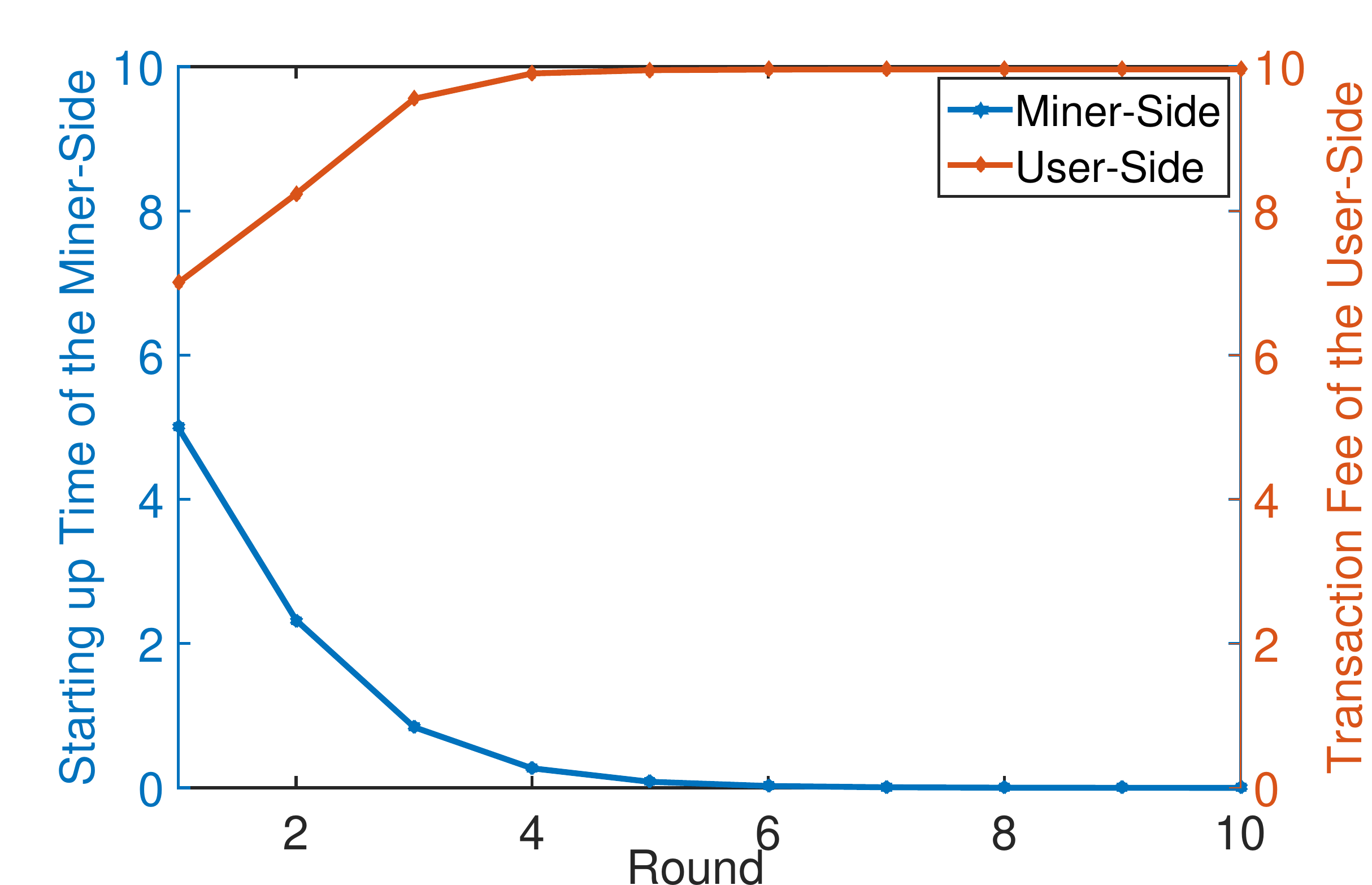}
\end{minipage}%
}%

\subfigure[$\mathbf{p^0}=\mathbf{0.7},\mathbf{q^0}=\mathbf{0.9}$.]{
\begin{minipage}[t]{0.47\linewidth}
\centering
\includegraphics[height=1.3in,width=1.75in]{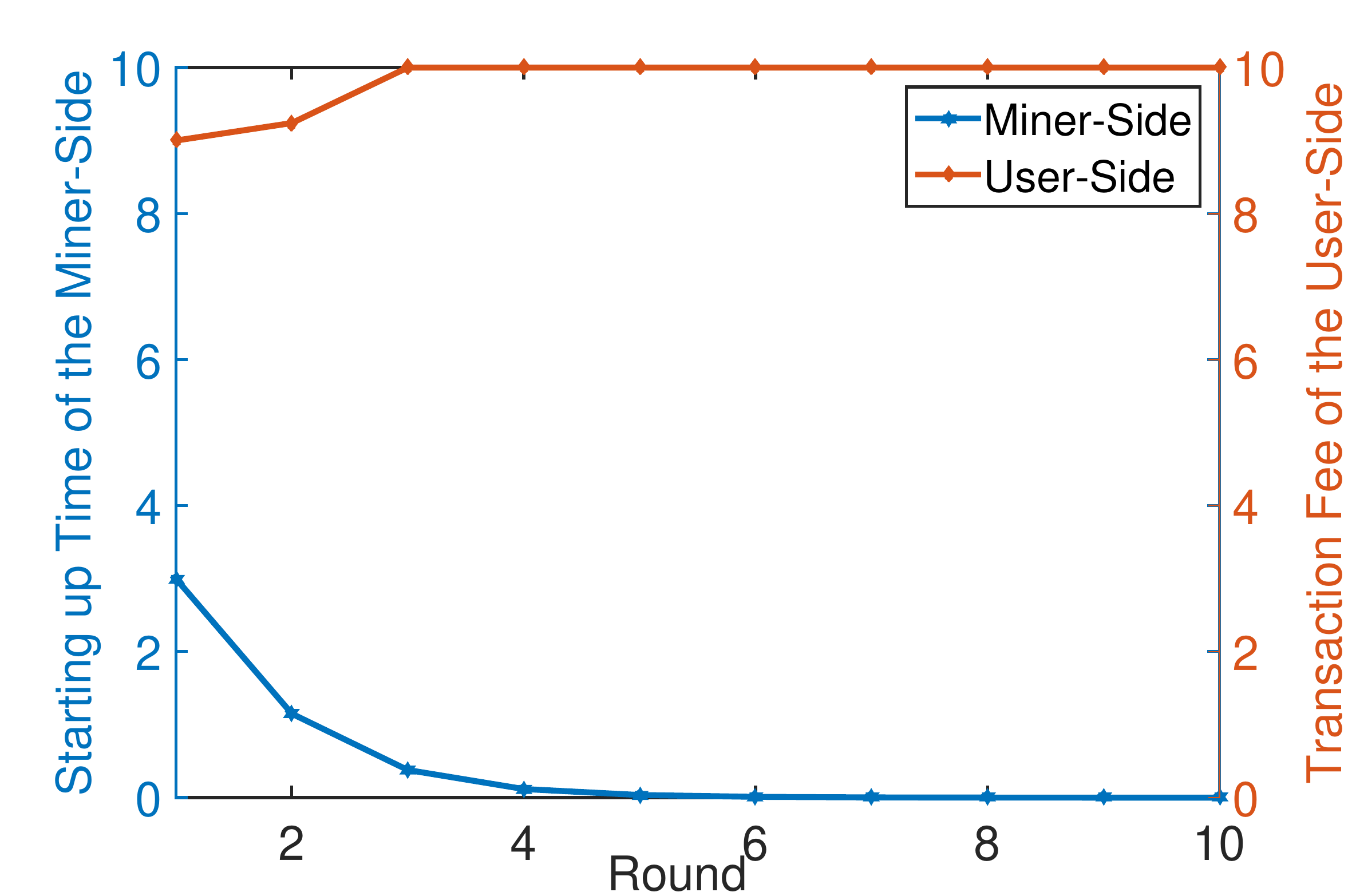}
\end{minipage}
}%
\subfigure[$\mathbf{p^0}=\mathbf{0.9},\mathbf{q^0}=\mathbf{0.3}$.]{
\begin{minipage}[t]{0.47\linewidth}
\centering
\includegraphics[height=1.3in,width=1.70in]{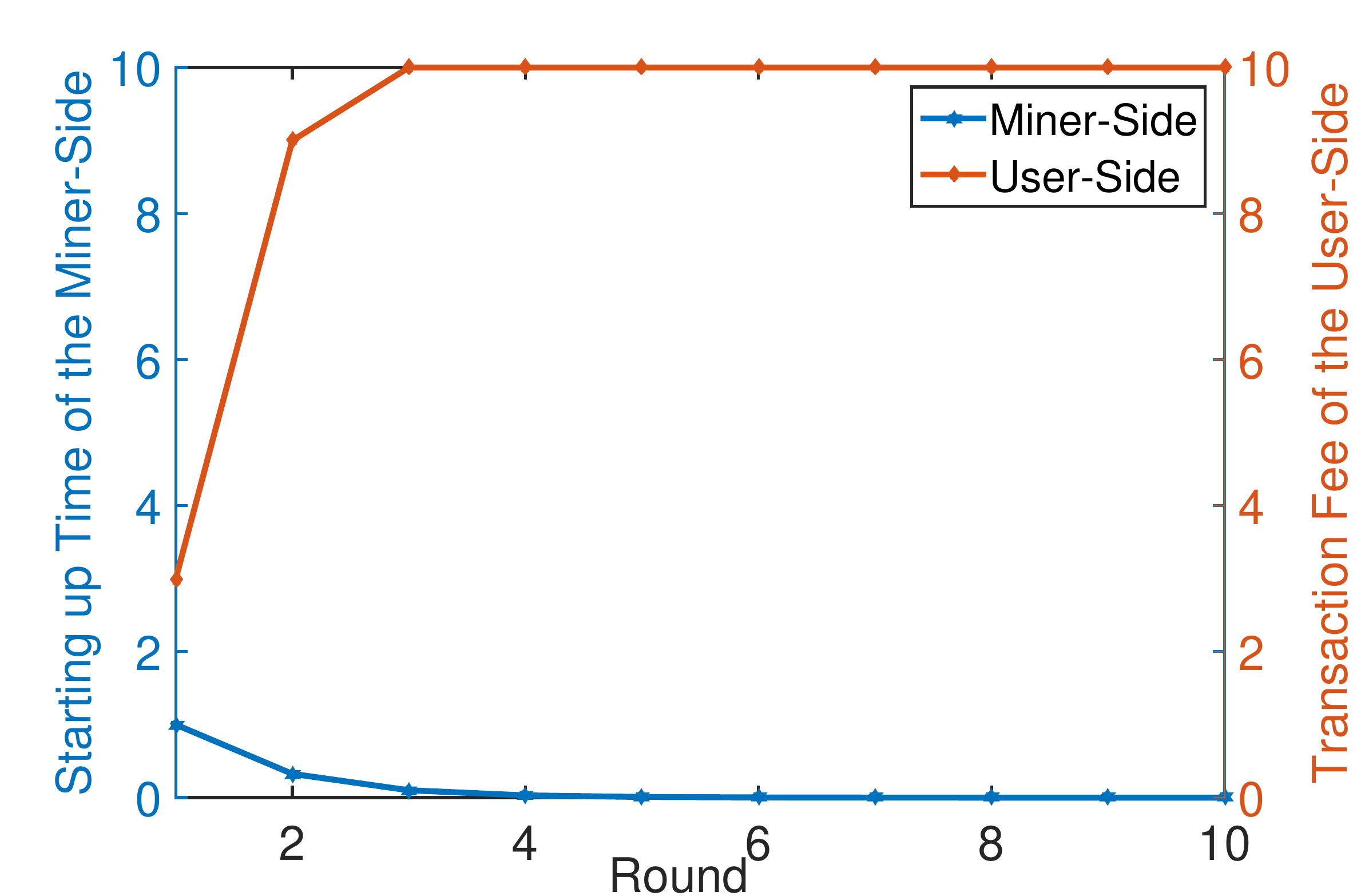}
\end{minipage}
}%
\centering
\caption{The strategy evolutions of the miner-side and user-side vs. various initial probability settings.}
\label{fig5}
\end{figure}

In this section, we testify the effectiveness of the proposed ZD-based incentive mechanism experimentally. To begin with, suppose
the profit functions and cost functions of the miner-side and user-side are defined to satisfy the linear relationship for simplicity with $\chi_m(y)=y+6.25$ because current Bitcoin subsidy is set as $\$6.25$,  $\chi_u(x)=x$, $\Xi_m(x)=x$, and $\Xi_u(y)=y$. And the scaling parameters are $\varpi_m=\varpi_u=0.4$, $\varkappa_m=\varkappa_u=0.6$ and $\omega_1=0.4,$ $\omega_2=0.8$.  Notably, other parameter settings and multiple monotonically increasing functions have been tested and derive very similar results, thus we omit to present them to avoid redundancy. Further, we partition the continuous strategy spaces $X=[0,1]$ and $Y=[0,10]$ into 10 sub-spaces in order to present a precise statistical calculation of the miner-side's state transition probability matrix $\{P_{a\rightarrow r}\}_{\eta_1\times\eta_1}$ with $T=10$. Note that the number of the preliminary phase is set as $R=100$ and we carry out $Q=200$ rounds for the simulation process. However, we only depict the game results of the first several rounds to get a clearer observation of the experimental trends. Each simulation is repeated 50 times so as to gain the average value for statistical confidence.

Fig. \ref{fig2} shows the evolutions of the earliest starting up probabilities of the miner-side when the user-side respectively adopts the proposed ZD-based incentive mechanism and five classical strategies, i.e., all cooperation (ALL-C), all defection (ALL-D), win-stay-lose-shift (WSLS), tit-for-tat (TFT) and random  strategy. In ALL-C (ALL-D) strategy, no matter what strategy employed last round, the user-side chooses the highest (lowest) bidding strategy for each round; in WSLS strategy, the user-side keeps the same action as the previous one if it creates a beneficial payoff, otherwise, it chooses the opposite strategy; in TFT strategy, the user-side selects the opposite action to that of the miner-side previously; in random strategy, the user-side sets its strategy arbitrarily for each round. We set the initial probabilities that the miner-side chooses the earliest starting up strategy and the user-side offering the maximal transaction fee as $\mathbf{p^0}=[p_{0-\eta_1}^0,...,p_{\eta_2-\eta_1}^0]=\mathbf{0.3}, \mathbf{0.5},\mathbf{0.7},\mathbf{0.9}$ and $\mathbf{q^0}=[q_{00-\eta_2}^0,...,q_{\eta_10-\eta_2}^0,...,q_{\eta_1\eta_2-\eta_2}^0]=\mathbf{0.5},\mathbf{0.7}, \mathbf{0.9}, \mathbf{0.3}$, respectively. Here, $\mathbf{0.3}$ denotes an 11-dimensional vector with all components 0.3, so does for $\mathbf{0.5}, \mathbf{0.7}, \mathbf{0.9}$. One can conclude that the probability of choosing the earliest strategy of the miner-side can always reach $1$ when the user-side employs the ZD-based incentive mechanism no matter what initial probabilities are set. However, other classical strategies fail. Such a superiority undoubtedly demonstrates the effectiveness of our ZD-based incentive mechanism in driving the miner-side to be an ``early bird".

Fig. \ref{fig4} displays the change in the probability of powering on rigs at the earliest time (Fig. \ref{fig4} (a)) and the starting up time (Fig. \ref{fig4} (b)) of the miner-side under different initial probability settings when fighting with the user-side who adopts the ZD strategy. From this, we can state that 1) the probability of choosing the earliest starting up strategy will eventually tend to 1; and 2) the starting up time of the miner-side decreases continuously until $t=0$, both implying that the miner-side is motivated to mine at the very beginning of each round finally. This result more directly validates the efficacy of our mechanism.

Fig. \ref{fig5} demonstrates that when the starting up time of the miner-side decreases with the help of the ZD strategy, the user-side is also confined to offer the highest transaction fee. This is an indication of the fairness of our ZD-based incentive mechanism, in which the ZD adopter has no way to squeeze the miner-side financially even can dominate the game.

\section{Conclusion}\label{conclusion}
In this work, we design a ZD-based incentive mechanism to cap the mining gap and address the egoistic dilemma presented in the transaction fee-incentive Bitcoin. We start by modeling the in-circle games among homogenous players as supermodular games and derive the strategic complementarity consequently. Based on this, the multi-miner and multi-user game can be simplified as a miner-side and user-side game. In such a game, we devise a powerful ZD-based incentive mechanism for the user-side to coerce the miner-side to be an ``early bird", making both players get rid of the egoistic dilemma successfully. Our mechanism has the {\it sustained ability of motivation} and is featured by {\it fairness}, showing its vitality over the long run.

\section*{Acknowledgment}

This work has been supported by National Key R\&D Program of China (No. 2019YFB2102600), National Natural Science Foundation of China (No. 61772080, 61672321, 61771289, 61832012, and 62072044), the Blockchain Core Technology Strategic Research Program of Ministry of Education of China (No. 2020KJ010301), BNU Interdisciplinary Research Foundation for the First-Year Doctoral Candidates (No. BNUXKJC2022), the International Joint Research Project of Faculty of Education, Beijing Normal University, and Engineering Research Center of Intelligent Technology and Educational Application, Ministry of Education.

\bibliographystyle{IEEEtran}
\bibliography{2020-12-21}

\begin{IEEEbiography}[{\includegraphics[width=1in,height=1.25in,clip,keepaspectratio]{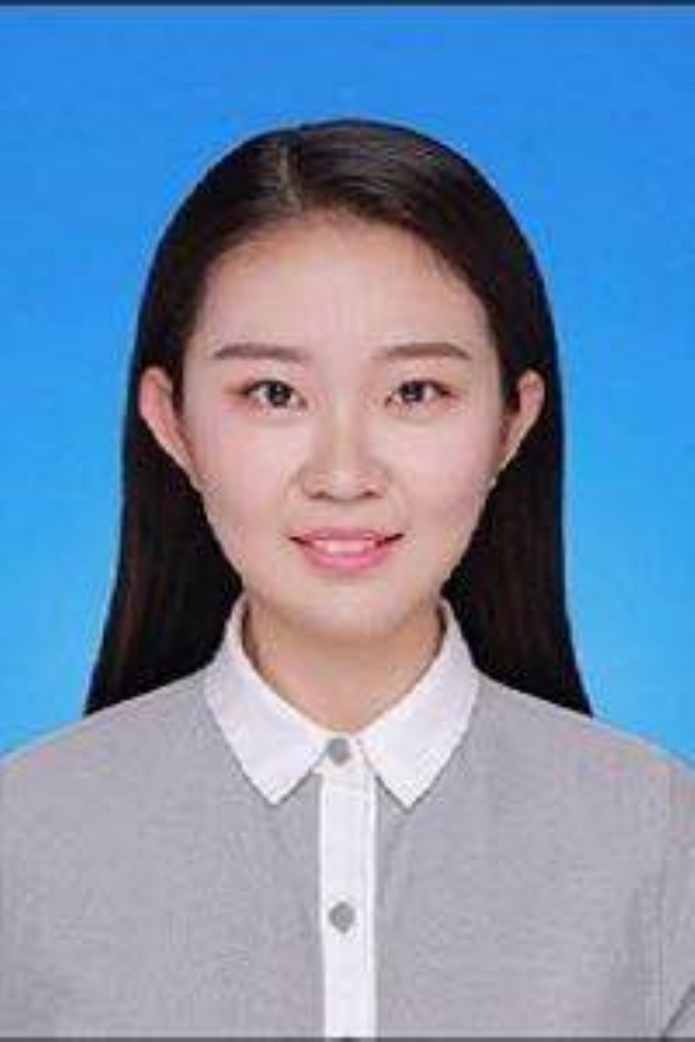}}]
{Hongwei Shi}received her B.S. degree in Computer Science from Beijing Normal University in 2018. Now she is pursuing her Ph.D. degree in Computer Science from Beijing Normal University. Her research interests include blockchain, game theory and combinatorial optimization.
\end{IEEEbiography}

\begin{IEEEbiography}[{\includegraphics[width=1in,height=1.25in,clip,keepaspectratio]{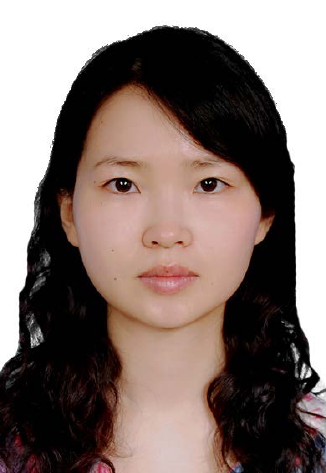}}]{Shengling Wang} is a full professor in the School of Artificial Intelligence, Beijing Normal University. She received her Ph.D. in 2008 from Xi an Jiaotong University. After that, she did her postdoctoral research in the Department of Computer Science and Technology, Tsinghua University. Then she worked as an assistant and associate professor from 2010 to 2013 in the Institute of Computing Technology of the Chinese Academy of Sciences. Her research interests include mobile/wireless networks, game theory, crowdsourcing.
\end{IEEEbiography}

\begin{IEEEbiography}[{\includegraphics[width=1in,height=1.25in,clip,keepaspectratio]{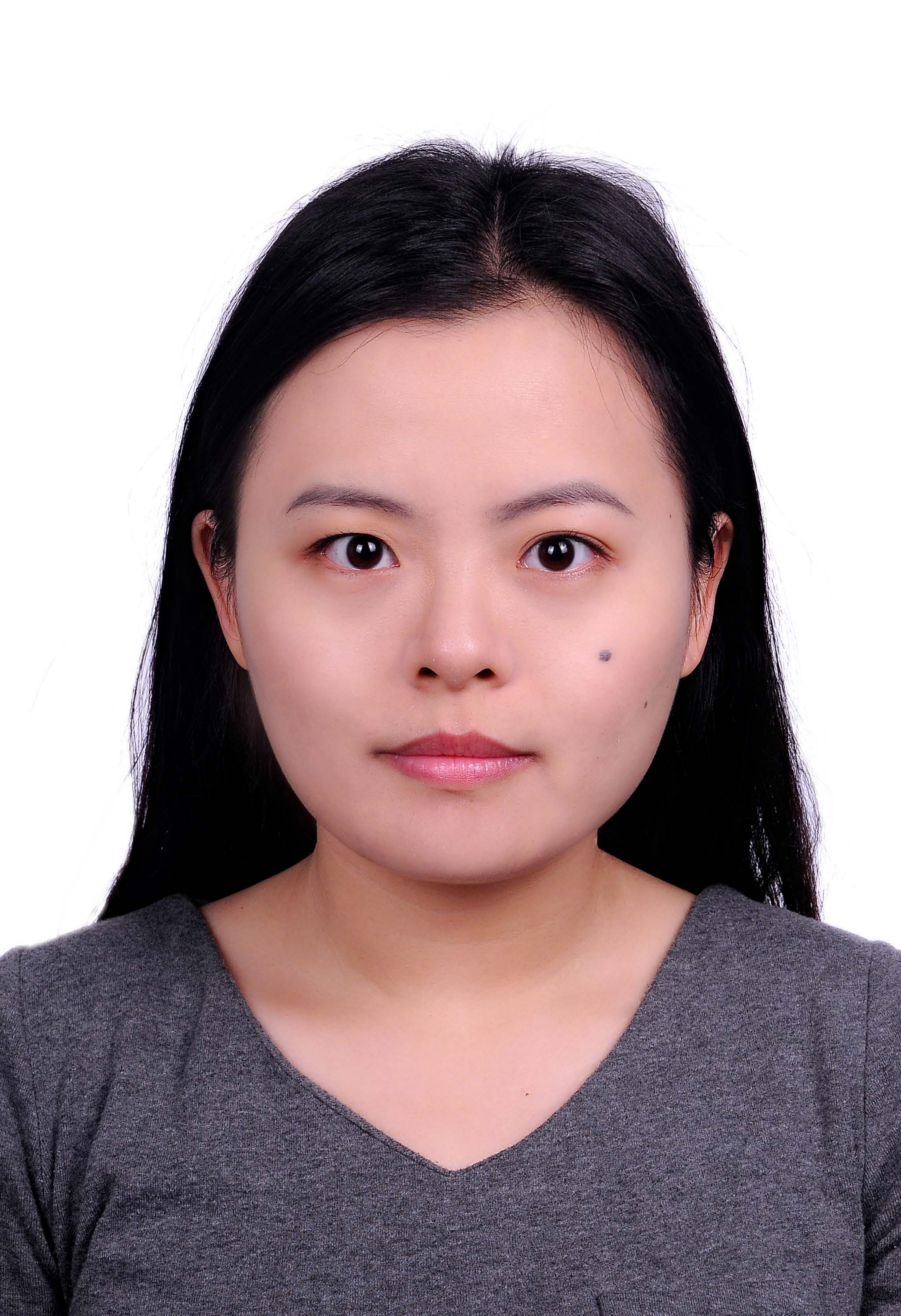}}]
{Qin Hu} received her Ph.D. degree in Computer Science from the George Washington University in 2019. She is currently an Assistant Professor in the department of Computer and Information Science, Indiana University - Purdue University Indianapolis. Her research interests include wireless and mobile security, crowdsourcing/crowdsensing and blockchain.
\end{IEEEbiography}

\begin{IEEEbiography}[{\includegraphics[width=1in,height=1.25in,clip,keepaspectratio]{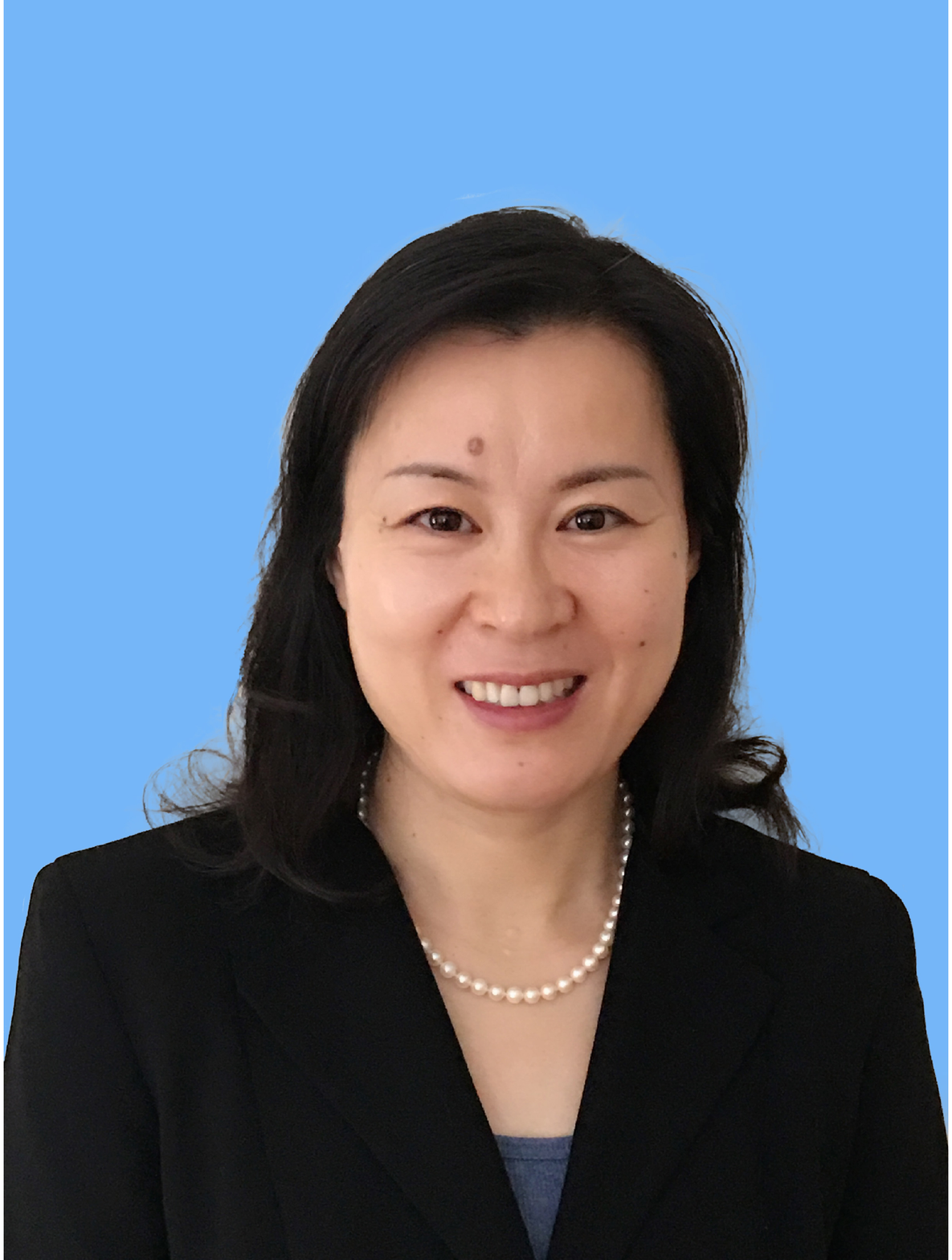}}]
{Xiuzhen Cheng}[F] received her M.S. and Ph.D. degrees in computer science from the University of Minnesota Twin Cities in 2000 and 2002. She is a professor in the Department of Computer Science, George Washington University, Washington, DC. Her current research interests focus on privacy-aware computing, wireless and mobile security, dynamic spectrum access, mobile handset networking systems (mobile health and safety), cognitive radio networks, and algorithm design and analysis. She has served on the Editorial Boards of several technical publications and the Technical Program Committees of various professional conferences/workshops. She has also chaired several international conferences. She worked as a program director for the U.S. National Science Foundation (NSF) from April to October 2006 (full time), and from April 2008 to May 2010 (part time). She published more than 170 peer-reviewed papers.
\end{IEEEbiography}

\begin{IEEEbiography}[{\includegraphics[width=1in,height=1.25in,clip,keepaspectratio]{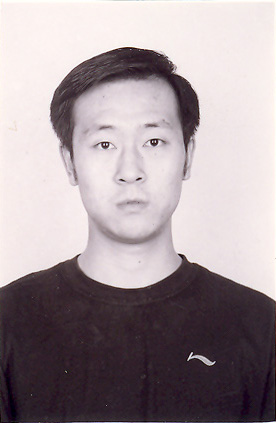}}]
{Jianhui Huang} received the Ph.D. degree in computer science from Xi an Jiaotong University, Xi an, China, in 2009. He is currently an Associate Professor with the Institute of Computing Technology, Chinese Academy of Sciences, Beijing, China. His current research interests include the mobile applications, opportunistic network, and cloud computing.
\end{IEEEbiography}

\end{document}